\documentclass[a4paper,11pt]{article}

\usepackage[fleqn]{amsmath}
\usepackage{slashed,graphicx,ulem,geometry,setspace,algorithm,algorithmic}
\usepackage{mathrsfs,pgffor,amssymb,fix-cm,bm,ulem,extarrows}
\usepackage[greek,english]{babel}
\usepackage{float,cite,color,subfigure}
\usepackage[figuresright]{rotating}
\usepackage{dsfont}               
\usepackage{mathrsfs}             
\usepackage{verbatim}
\usepackage{slashed,graphicx,epsfig,ulem,geometry,setspace,algorithm,algorithmic,mathrsfs,pgffor,amssymb,fix-cm,extarrows}
\def\eq{\begin{eqnarray}}
\def\en{\end{eqnarray}}
\def\TT{\tilde T}
\newcommand{\Red}{\textcolor{red}}

\renewcommand\sout{\bgroup \color{red} \ULdepth=-.5ex \ULset}

\definecolor{orange}{RGB}{255,127,0}
\definecolor{pink}{RGB}{255,192,203}
\definecolor{brown}{RGB}{139,35,35}
\definecolor{Magenta}{RGB}{255,0,255}
\begin{document}
\date{}
\title{Electromagnetic and gravitational form factors of $\Delta$ resonance in a covariant quark-diquark approach}
\author{Dongyan Fu$^{a,b,\thanks{fudongyan@ihep.ac.cn}}$, Bao-Dong Sun$^{c,d,\thanks{bao-dong.sun@m.scnu.edu.cn}}$,
and Yubing Dong$^{a,b,\thanks{dongyb@ihep.ac.cn}}$\\
Institute of High Energy Physics, Chinese Academy of Sciences, Beijing 100049, China$^{a}$\\
School of Physical Sciences, University of Chinese Academy of Sciences, Beijing 101408, China$^{b}$\\
Guangdong Provincial Key Laboratory of Nuclear Science,
Institute of Quantum Matter, \\South China Normal University, Guangzhou 510006, China$^{c}$\\
Guangdong-Hong Kong Joint Laboratory of Quantum Matter,\\
Southern Nuclear Science Computing Center, \\South China Normal University, Guangzhou 510006, China$^{d}$}
\newenvironment{mysubeq}
{\begin{subequations}
\renewcommand\theequation{\theparentequation \alph{equation}}}
{\end{subequations}}
\vspace{1em}
\maketitle

\begin{abstract}
In this work, the electromagnetic and gravitational form factors of a spin-$3/2$ particle, $\Delta$ resonance, are simultaneously calculated with the
help of a relativistic covariant quark-diquark approach. The two kinds of form factors are separately extracted from the matrix elements of the
electromagnetic current and of the energy-momentum tensor of the system. Our numerical results show that the approach can well reproduce the
electromagnetic monopole, dipole, quadrupole, and octupole form factors comparing to the Lattice calculations. Our obtained electromagnetic moments
are also comparable with some other approaches. Moreover, the obtained gravitational form factors, which give the mechanical properties of the system
like the mass and spin distributions, are also displayed for the $\Delta$ isobar. In addition, some discussions of the sign and the interpretation
of the D-term are particularly given.\\
\end{abstract}
\begin{small}
\vspace{0.5cm}
 \par
{\hskip 1.5cm}$D$-term; Quark-diquark approach.
\end{small}
\vspace{0.5cm}
\section{Introduction}
\par\noindent\par
It is well known that the electromagnetic form factors (EMFFs) are the indispensable physical quantities in revealing the internal structure of a
complicated system. The electromagnetic form factors of hadrons, like $\pi$-meson and nucleon, can tell the charge or magnetic distributions
of the systems. They also illustrate the charge and magnetic radii, which can be extracted by the slopes of the charge and magnetic distributions
of the systems at $q^2=0$ (with $q$ being the momentum transfer)~\cite{Gross:2006fg,deMelo:2008rj,Cloet:2014rja,deAraujo:2017uad}. Furthermore, for a
spin-1 system, for instance a deuteron or a vector meson of $\rho$, its charge, magnetic, and quadrupole form factors can embody its intrinsic structures
as well, such as its charge and magnetic distributions and quadrupole deformation, (see
Refs.~\cite{Gross:2002ge,Gross:2001ap,Gilman:2001yh,Garcon:2001sz,Sun:2016ncc,Dong:2008mt}
for the deuteron, and Refs.~\cite{Krutov:2018mbu,Krutov:2016uhy,Choi:2004ww,deMelo:1997hh,Sun:2017gtz} for
the $\rho$ meson, respectively). Consequently, EMFFs can provide discriminating information for studying the inner structures of hadrons.\\

There are many studies devoted to the understanding of the electromagnetic form factors of the nucleon, its excitations $N^*$, and the well-known
$N-\Delta$ transitions in the literature. The constituent quark model is one of the successful approaches. In those quark model calculations,
the nucleon or its excitation is regarded as a three quark system and the electromagnetic current probes each quark
~\cite{Capstick:1989ck,Capstick:1992uc,Capstick:1992xn,Giannini:2002tn,Lipes:1972nf} instantaneously. Then, the form factors are obtained by
the calculation of the three quark contributions to the matrix element by using the wave function of the hadron. Relativistic corrections to
the wave function of the nucleon or its excitations, as well as to the electromagnetic interaction operator may also be taken into account in those
quantum mechanical calculations. Reasonable results comparing to the experimental measurement can be obtained. It should be mentioned that, different
from those calculations (with some relativistic corrections), the relativistic covariant quark-diquark approach is
also employed to study the electromagnetic form factors of nucleon~\cite{Kroll:1991ag,Keiner:1995bu,Meyer:1994cn,Nagata:2005ve,Keiner:1996at,Zhang:2016qqg}.
In those relativistic covariant field theory studies, the diquark contribution, as well as the quark one, are simultaneously and explicitly
considered. Their results are also well consistent to the available experimental data.\\

Besides the electromagnetic form factors of hadrons, the gravitational form factors (GFFs) are also expected to embody the fundamental information
of the spatial distributions, like the energy, spin, and strong forces~\cite{Polyakov:2002yz} of systems. Those GFFs are defined through the
matrix element of the symmetric energy-momentum tensor (EMT). More details about GFFs can be found in
Refs.~\cite{Diehl:2003ny,Belitsky:2005qn,Polyakov:2018exb,Polyakov:2018zvc,Polyakov:2002yz,Lorce:2018egm}. Clearly, GFFs describe the interaction
between the gravitation, as an external field, and the matter fields, in which the scattering off the graviton is a natural but impractical probe
for GFFs. Luckily, because of the similar structure of EMT and electromagnetic current operators~\cite{Diehl:2003ny}, hard-exclusive reactions,
like deeply virtual Compton scattering (DVCS) and vector meson electro-production, provide a realistic way to access the GFFs of hadrons through the
generalized parton distributions (GPDs)~\cite{Polyakov:2019lbq,Sun:2020wfo} and through the generalized distribution amplitudes
(GDAs)~\cite{Kumano:2017lhr}. It is expected that the nucleon GPDs will be measured at some facilities, such as Jefferson Lab (JLab.), the future
Electron-Ion Collider (EIC)~\cite{Proceedings:2020eah}, and the one in China(EicC)~\cite{Chen:2018wyz}.\\

One reason why GFFs are extremely important partially comes from their connections to the (especially) GPDs and GDAs. It is believed that GPDs are an
important metric of the three-dimensional hadron structure, and they can be loosely described as amplitudes for removing a parton from a hadron and
replacing it with one with different momentum. In addition, the moments of GPDs are related not only to EMFFs but also to GFFs, and
one of the GFFs describes the total angular momentum carried by the partons. It is regarded that the finding the contribution to the sum of the spin
and orbital angular momenta from specific components of hadrons is of great importance~\cite{Ji:1996ek,Goeke:2001tz,Diehl:2003ny,Belitsky:2005qn}.
In particular, it is discussed that there is a very important quantity $D$-term~\cite{Polyakov:1999gs}, which closely relates to the matrix element
of EMT $T^{ij}$ components. As the energy and angular momentum, the $D$-term is also corresponding to the values of GFFs at zero momentum transfer.
Therefore, the $D$-term is considered as the "last global unknown property", which is believed to characterize the spatial deformations as well as other mechanical properties of hadrons~\cite{Polyakov:2002yz}.\\

For the GFFs of hadrons with spin 0, 1/2, and 1, much work has been already done~\cite{Pagels:1966zza,Kumano:2017lhr,Kim:2012ts, Azizi:2019ytx,
Cosyn:2018thq,Broniowski:2008hx,Freese:2021mzg}. The common approaches of the chiral quark model, LQCD
calculation, the effective chiral theory, the SU(2) skyrme model, the bag model, the QCD sum rule, and the AdS/CFT correspondence
~\cite{Freese:2019eww,Broniowski:2008hx,Kim:2020nug,Shanahan:2018pib,Alharazin:2020yjv,Kim:2012ts,Neubelt:2019sou,Azizi:2019ytx,Hatta:2018ina} have all
been employed. Although there are some approaches devoted for the GFFs of a spin-3/2 $\Delta$ resonance~\cite{Kim:2020lrs,Panteleeva:2020ejw,Pefkou:2021fni},
simultaneous discussion and calculation of the EMFFs and GFFs for $\Delta$ is still missing. Different from the hadrons with spin-0, 1/2, and 1,
$\Delta(1232)$ is a low-lying baryon resonance with spin-3/2, the study of its EMFFs and GFFs can give more information about the internal structures
of this high-spin particle and can be further applied for the transition EMFFs and GFFs of $N-\Delta$ process~\cite{Pascalutsa:2006ne}. Therefore,
such a study is of great interest. In this work, we employ the relativistic and covariant quark-diquark approach to simultaneously calculate the
EMFFs and GFFs of the spin-3/2 $\Delta$ particle. We know that the baryon $\Delta$ can be simply regarded as a three quark system, and here
we treat it as a system of a quark plus an axial vector diquark. Consequently, the estimated form factors are given by the sum of
quark and diquark contributions. \\

This paper is organized as follows. In section~\ref{section2}, the definitions of EMFFs and GFFs for a spin-3/2 particle are given. Section~\ref{section3} shows the
corresponding matrix elements of the quark and diquark for the electromagnetic and gravitational probes in the covariant quark-diquark approach.
In section~\ref{section4}, the model parameters are firstly determined comparing to the Lattice calculations for the EMFFs of $\Delta^{+}$. Then, our
numerical calculations for the electric monopole, magnetic dipole, electric quadrupole, and magnetic octupole form factors are given.
Finally, we display our calculated GFFs of $\Delta$, such as its mass and spin distributions, and we particularly address the issues of the sign and the interpretation
of the $D$-term. Section~\ref{section5} is devoted to a summary.\\

\section{Form factors of a spin-3/2 particle}\label{section2}
\subsection{Electromagnetic Form Factors}\label{sub2_1}
\par\noindent\par
It is well known that in the one-photon approximation a composite particle with spin-$S$ has $(2S+1)$ independent electromagnetic form factors due to
the symmetries and conservations, like parity and time-reversal. For the spin-3/2 particle, the matrix element of the electromagnetic current is expressed as~\cite{Cotogno:2019vjb}
\begin{equation}\label{vectorcurrent}
    \begin{split}
    \left\langle p^\prime,\lambda^\prime \left| \hat{J}_a^{\mu}\left( 0 \right)
    \right| p,\lambda\right\rangle=&-\bar{u}_{\alpha'} \left( p',\lambda' \right)
    \biggl[ \frac{P^\mu}{M} \left( g^{\alpha' \alpha } F_{1,0}^{V,a} \left( t \right)
    -\frac{q ^{\alpha'} q ^\alpha}{2M^2} F_{1,1}^{V,a} \left( t \right)\right)\\
    &+\frac{i \sigma^{\mu \nu} q_\nu}{2M} \left( g^{\alpha' \alpha} F_{2,0}^{V,a} \left( t \right)
    -\frac{q ^{\alpha'} q ^\alpha}{2M^2}F_{2,1}^{V,a} \left( t \right)\right) \biggr]
    u_\alpha \left( p,\lambda \right),
    \end{split}
\end{equation}
where $u_\alpha \left( p,\lambda\right)$ is the known Rarita-Schwinger spinor for a spin-3/2 particle. In general, the index $a$ in
Eq.~\eqref{vectorcurrent} runs from a gluon to quark flavors and the total form factors $F_{i,j} = \Sigma_a F^{V,a}_{i,j}$. In precent work, we only consider the constituent
quark (and diquark) degrees of freedom and do not take the gluon contribution into account. In this work,
we introduce the kinematical variables $P^\mu = (p^\mu + p'^\mu)/2$, $q^\mu = p'^\mu-p^\mu$, and $q^2=t$ (which stands for the squared momentum
transfer), where $p(p')$ is the initial (final) momentum. The normalization of the Rarita-Schwinger spinor is taken to be $\bar{u}_{\sigma'}(p)u_\sigma
(p)= -2M\delta_{\sigma' \sigma}$.\\

In the non-relativistic approximation, the EMFFs can be further expressed in terms of $F^V_{i,0(1)}~(i=1,2)$
(according to Eq.~\eqref{vectorcurrent} and Ref.~\cite{Nozawa:1990gt}). In the Breit frame, the average of the baryon momenta and the momentum
transfer are respectively defined by $P^\mu=(E,0,0,0)$ and $q^\mu = (0,\bm{q})$. Thus,
$q^2 = - \bm{q}^2 =t =4(M^2-E^2)$ with the $\Delta$ isobar mass being $M$. Then,
\begin{mysubeq}\label{EMFFsequation}
    \begin{align}
        G_{E0}\left( t \right)=&\left( 1+\frac{2}{3}\tau \right) [F_{2,0}^V(t) + (1+\tau)(F_{1,0}^V(t)-F_{2,0}^V(t))]\nonumber
    \\
        & +\frac{2}{3} \tau (1+\tau) [F_{2,1}^V(t) + (1+\tau)(F_{1,1}^V(t)-F_{2,1}^V(t))],
    \\
        G_{E2}\left( t \right)=& [F_{2,0}^V(t) + (1+\tau)(F_{1,0}^V(t)-F_{2,0}^V(t))] + (1+\tau) [F_{2,1}^V(t) + (1+\tau)(F_{1,1}^V(t)-F_{2,1}^V(t))],
    \\
        G_{M1}\left( t \right)=&\left(1+\frac{4}{5}\tau\right) F_{2,0}^V \left( t \right)
        +\frac{4}{5} \tau  (\tau +1) F_{2,1}^V \left( t \right),
    \\
        G_{M3}\left( t \right)=& F_{2,0}^V \left( t \right)
        + (\tau +1) F_{2,1}^V \left( t \right),
    \end{align}
\end{mysubeq}
\noindent{\hskip -0.35cm}
where $\tau = -t/4M^2~(\ge 0)$. In Eq.~\eqref{EMFFsequation}, $G_{E0}$, $G_{E2}$, $G_{M1}$, and $G_{M3}$ are the charge,
electric-quadrupole, magnetic-dipole and magnetic-octupole form factors, respectively. When the momentum transfer square goes to zero,
namely $t=-q^2\to 0$, we get the charge, magnetic-dipole, electric-quadrupole, and magnetic-octupole moments. Moreover, the slope of electric
monopole form factor shows the corresponding charge radius of the system. According to Ref.~\cite{Leinweber:1992hy}, we have
\begin{equation}\label{rE}
    \left\langle r^2 \right\rangle_E = 6 \frac{d}{dt}\tilde{G}_{E0}(t) \Big \vert _{t=0},
\end{equation}
in which $\tilde{G}_{E0}$ has been normalized $\tilde{G}_{E0}=\frac{G_{E0}}{Q_e}$, and $Q_e$ is the charge quantum number carried by the particle.\\

\subsection{Gravitational Form Factors}\label{sub2_2}
\par\noindent\par
The GFFs for a spin-3/2 particle are defined through the matrix element of its EMT tensor as~\cite{Cotogno:2019vjb,Kim:2020lrs}
\begin{equation}\label{GFFsSpin3/2}
    \begin{split}
    &\left\langle p^\prime,\lambda^\prime \left| \hat{T}^{\mu \nu}(0)
    \right| p,\lambda\right\rangle  \\
    &~~~=-\bar{u}_{\alpha ^\prime}\left(p^\prime,\lambda^\prime\right) \bigg [\frac{P^\mu P^\nu}{M} \left(g^{\alpha'  \alpha} F_{1,0}^T (t)
    -\frac{q ^{\alpha' } q ^{\alpha}}{2 M^2}F_{1,1}^T (t) \right)
    + \frac{ \left({q }^\mu {q }^\nu- {g}^{\mu  \nu } q ^2\right) }{4M}
    \left({g}^{\alpha'  \alpha}F_{2,0}^T (t)-\frac{{q }^{\alpha '}
    {q }^{\alpha}}{2 M^2}F_{2,1}^T (t)\right)  \\
    &~~~~~+ M g^{\mu  \nu } \left(
    g^{\alpha'  \alpha}F_{3,0}^T (t)-\frac{ q^{\alpha'}
    q^{\alpha}}{2 M^2}F_{3,1}^T (t)\right)
    + \frac{i {P}^{ \{ \mu } \sigma ^{\nu \} \rho} q_\rho}{2M} \left(g^{\alpha'  \alpha}F_{4,0}^T (t)
    -\frac{q ^{\alpha' } q ^{\alpha}}{2 M^2}F_{4,1}^T (t)\right) \\
    &~~~~~- \frac{1}{M} \left({q }^{\{ \mu} g^{\nu \} \{ \alpha '} {q }^{\alpha \}}
    -2 q^{\alpha' } q^{\alpha} g^{\mu  \nu }
    - g^{\alpha ' \{ \mu } g^{\nu \} \alpha } q^2 \right) F_{5,0}^T (t)
    + M g^{\alpha ' \{ \mu } g^{\nu \} \alpha}F_{6,0}^T (t) \bigg ]
    u_\alpha \left( p,\lambda\right).
    \end{split}
\end{equation}
The above definition is for the total GFFs of the system. One can also define the contributions of the quark and gluon individually.
Here, $F_{3,0}^T, F_{3,1}^T$ and $F_{6,0}^T$ are nonconserving terms, and they should be vanishing if we consider the total EMT. Since only the
quark contributions is taken into account in our present approach, $F_{3,0}^T, F_{3,1}^T$ and $F_{6,0}^T$ are simply ignored. Moreover, the convention
$a^{\{ \mu}b^{\nu \}} = a^\mu b^\nu + a^\nu b^\mu$ and $a^{ [ \mu}b^{\nu ] } = a^\mu b^\nu - a^\nu b^\mu$ are adopted. \\

In the Breit frame, the gravitational multipole form factors (GMFFs) are derived from the matrix element of the EMT current~\cite{Cotogno:2019vjb,Kim:2020lrs}. Here we summarize the results as followings.
\begin{align}
\langle p', \sigma' | \hat{T}^{00}(0) | p , \sigma \rangle &= 2ME \bigg{[} \varepsilon_{0}(t)\delta_{\sigma'\sigma}
+ \left(\frac{\sqrt{-t}}{M}\right)^{2} \hat{Q}_{\sigma'\sigma}^{kl}Y^{kl}_{2} \varepsilon_{2}(t)\bigg{]}, \cr
\langle p', \sigma' | \hat{T}^{0i}(0) | p , \sigma \rangle &= 2ME\bigg{[}
\frac{\sqrt{-t}}{3M}i\epsilon^{ikl}Y_{1}^{l}\hat{S}^{k}_{\sigma'\sigma}\mathcal{J}_{1}(t)
+\left(\frac{\sqrt{-t}}{M}\right)^{3}i\epsilon^{ikl}Y^{lmn}_{3}\hat{O}^{kmn}_{\sigma'\sigma}\mathcal{J}_{3}(t)\bigg{]}, \cr
\langle p', \sigma' | \hat{T}^{ij}(0) | p , \sigma \rangle &=2ME\bigg{[} \frac{1}{4M^{2}}(\Delta^{i}\Delta^{j} + \delta^{ij}
\Delta^{2}) D_{0}(t)\delta_{\sigma'\sigma }+ \frac{1}{4M^{4}} \hat{Q}^{kl}_{\sigma'\sigma}(\Delta^{i}\Delta^{j} + \delta^{ij}\Delta^{2}) \Delta^{k}
\Delta^{l}D_{3}(t) \cr
&\hspace{1cm}+ \frac{1}{2M^{2}} \left(\hat{Q}^{ik}_{\sigma'\sigma} \Delta^{j}\Delta^{k} +\hat{Q}^{jk}_{\sigma'\sigma}  \Delta^{i}\Delta^{k}
  +  \hat{Q}^{ij}_{\sigma'\sigma} \Delta^{2} - \delta^{ij}\hat{Q}^{kl}_{\sigma'\sigma} \Delta^{k}\Delta^{l} \right)D_{2}(t)  \bigg{]},
  \label{eq:1}
\end{align}
where the spin-3/2 quadrupole- and octupole-spin operators $\hat{Q}^{ij}$ and $\hat{O}^{ijk}$ are respectively defined as
\begin{align}
\hat{Q}^{ij} &= \frac{1}{2}\left( \hat{S}^{i}\hat{S}^{j} +\hat{S}^{j}\hat{S}^{i} -\frac{2}{3}S(S+1)\delta^{ij}\right), \cr
\hat{O}^{ijk} &= \frac{1}{6}\bigg{(}
\hat{S}^{i}\hat{S}^{j}\hat{S}^{k}+\hat{S}^{j}\hat{S}^{i}\hat{S}^{k}+\hat{S}^{k}\hat{S}^{j}\hat{S}^{i}+\hat{S}^{j}\hat{S}^{k}
\hat{S}^{i}+\hat{S}^{i}\hat{S}^{k}\hat{S}^{j}+\hat{S}^{k}\hat{S}^{i}\hat{S}^{j}  \cr
&\hspace{0.3cm}-\frac{6S(S+1)-2}{5}(\delta^{ij}\hat{S}^{k}+\delta^{ik}\hat{S}^{j}+\delta^{kj}\hat{S}^{i})\bigg{)},
\end{align}
with $i,j,k=1,2,3$, and the spin operators can be expressed in terms of the SU(2) Clebsch-Gordan coefficients in the spherical basis as
\begin{align}
\hat{S}^{a}_{\sigma'\sigma} = \sqrt{S(S+1)}C^{S \sigma'}_{S \sigma 1 a} \ \ \ \mathrm{with} \ \ \ (a=0,\pm1. \  \  \sigma,\sigma'=0, \cdot\cdot\cdot,\pm S).
\end{align}
{\hskip -0.2cm}
Obviously, the GMFFs $\varepsilon_{0,1}(t)$, $\mathcal{J}_{0,1}(t)$ respectively relate to the matrix elements of $T^{00}$ and $T^{i0,0i}$,
and $D_{0,2,3}(t)$ to the ones of $T^{ij}$. They show the fundamental mechanical properties of the system. $\varepsilon_{0,1}(t)$, $\mathcal{J}_{0,1}(t)$
display the energy and angular momentum distributions, and $D_{0,2,3}(t)$ are interpreted as the essential quantities for characterizing the distributions of strong forces inside the system. \\

The relations among the GMFFs and GFFs are
\begin{mysubeq}\label{GMFFs_GFFs}
    \begin{align}
        {\varepsilon}_0 \left( t \right) & = F^T_{1,0}(t)  + \frac{t}{6 M^2}\biggl[
        - \frac{5}{2} F^T_{1,0}(t) - F^T_{1,1}(t) - \frac{3}{2} F^T_{2,0}(t) + 4 F^T_{5,0}(t) + 3 F^T_{4,0} \biggr]\notag\\
        &+ \frac{t^2}{12M^4} \biggl[ \frac{1}{2} F^T_{1,0}(t) + F^T_{1,1}(t) + \frac{1}{2} F^T_{2,0}(t)
        + \frac{1}{2} F^T_{2,1}(t) - 4 F^T_{5,0}(t) - F^T_{4,0}(t) - F^T_{4,1}(t) \biggr]\notag\\
        &+ \frac{t^3}{48M^6} \biggl[ - \frac{1}{2} F^T_{1,1}(t) - \frac{1}{2} F^T_{2,1}(t)
        + F^T_{4,1}(t) \biggr],\\
        {\varepsilon}_2(t) & = - \frac{1}{6} \biggl[ F^T_{1,0}(t) + F^T_{1,1}(t) -4 F^T_{5,0}(t)  \biggr]\notag\\
        & + \frac{t}{12M^2} \biggl[ \frac{1}{2} F^T_{1,0}(t) + F^T_{1,1}(t) + \frac{1}{2}F^T_{2,0}(t)
        + \frac{1}{2}F^T_{2,1}(t) - 4 F^T_{5,0}(t) -F^T_{4,0} -F^T_{4,1}(t)  \biggr]\notag\\
        & + \frac{t^2}{48M^4} \biggl[ -\frac{1}{2} F^T_{1,1}(t) - \frac{1}{2}F^T_{2,1}(t) + F^T_{4,1}(t) \biggr], \\
        \mathcal{J}_1(t) &= F^T_{4,0}(t)
        - \frac{t}{5M^2} \biggl[ F^T_{4,0}(t) + F^T_{4,1}(t) + 5F^T_{5,0}(t) \biggr]
        + \frac{t^2}{20M^4}F^T_{4,1}(t),\\
        \mathcal{J}_3(t) &= - \frac{1}{6} \biggl[ F^T_{4,0}(t) + F^T_{4,1}(t) \biggr] + \frac{t}{24M^2}F^T_{4,1}(t), \\
        D_{0}(t) &= F^T_{2,0}(t) - \frac{16}{3}F^T_{5,0}(t) -\frac{t}{6M^{2}}\bigg{[} F^T_{2,0}(t) +F^T_{2,1}(t) -4F^T_{5,0}(t)\bigg{]}
        +\frac{t^{2}}{24M^{4}}F^T_{2,1}(t),\\
        D_{2}(t) &=\frac{4}{3} F^T_{5,0}(t), \\
        D_{3}(t) &= \frac{1}{6}\bigg{[}-F^T_{2,0}(t) -F^T_{2,1}(t) + 4 F^T_{5,0}(t)\bigg{]}+\frac{t}{24M^{2}}F^T_{2,1}(t).
    \end{align}
\end{mysubeq}

One can also proceed by calculating the Fourier transformations of GMFFs to get the monopole and quadrupole densities~\cite{Kim:2020lrs}
\begin{equation}\label{epsi}
    \mathcal{E}_0(r) = M \widetilde{{\varepsilon}}_0(r), \qquad
    \mathcal{E}_2(r) = - \frac{1}{M} r \frac{d}{dr}\frac{1}{r} \frac{d}{dr}
    \widetilde{{\varepsilon}}_2(r),
\end{equation}
with $\widetilde{{\varepsilon}}_{0,2}(r) = \int \frac{d^3 q}{(2\pi)^3}
    e^{-i \bm{q} \cdot \bm{r}} {\varepsilon}_{0,2}(t)$ being the densities in $r$-space.\\

The mass radius of $\Delta$ is an important property, and it can be derived as~\cite{Kim:2020lrs}
\begin{equation}\label{rM}
    \left\langle r^2 \right\rangle_M = 6 \frac{d}{dt}\varepsilon_0(t) \vert _{t=0}.
\end{equation}

Moreover, if one interprets the static $T^{ij}(\mathbf{r})$ connecting to the pressure $p (r)$ and shear force $s (r)$ of the system like classical mechanics,
these two physical quantities relate to the $D$-term as~\cite{Polyakov:2018zvc}:
\begin{equation}
  \begin{split}
    p (r) & = \frac{1}{6 M} \frac{1}{r^2} \frac{d}{d r} r^2 \frac{d}{d r}
    \tilde{D}_0 (r),\\
    s (r) & = - \frac{1}{4 M} r \frac{d}{d r} \frac{1}{r} \frac{d}{d r}
    \tilde{D}_0 (r),\\
    \tilde{D}_0 (r) & = \int \frac{d^3 q}{(2 \pi)^3} e^{- i \bm{q} \cdot \bm{r}} D_0 (t).
  \end{split}
\end{equation}
According to Ref.~\cite{Perevalova:2016dln}, for the system the force on an infinitesimal piece
of area $d S^j$ at the distance $r$ has the form $F^i (\bm{r}) = T^{ij} (\bm{r}) d S^j = \left[ \frac{2}{3} s(r) + p(r) \right] dS^i$
where $dS^j = dS~r^j/r$. And the corresponding force must be directed outwards for the mechanical stability of the system.
Therefore the local criterion for the mechanical stability can be formulated
as~\cite{Perevalova:2016dln}
\begin{equation}
  p (r) + \frac{2}{3} s (r) ~>~ 0.
\end{equation}
Here we can express the $D$-term, $D = D_0 (0)$, by $p (r)$ and $s (r)$ as
\begin{equation}
  D = M \int d^3 r r^2 p (r) = - \frac{4 M}{15} \int d^3 r r^2 s (r) .
\end{equation}
So
\begin{equation}\label{Dtermps}
  M \int d^3 r r^2 p (r) + \frac{2 M}{3} \int d^3 r r^2 s (r) = - \frac{3}{2}
  D = M \int d^3 r r^2 \left( p (r) + \frac{2}{3} s (r) \right) ~>~0,
\end{equation}
which implies $D ~<~ 0$ for any stable system.

\vspace{0.5cm}

\section{Covariant quark-diquark approach}\label{section3}
\par\noindent\par
It is believed that the $\Delta$ isobar is composed of three light quarks, $u$ quark and $d$ quark. Since it has $I(J^p)=3/2(3/2^+)$, the total antisymmetry
makes its isospin and spin of each pair of quarks being 1. Here we treat two of them as a diquark. Therefore, the matrix element of the electromagnetic (or EM)
current is the sum of the contributions of the quark and diquark. For example, $\Delta^+$ contains two $u$ quarks and
one $d$ quark. So we can treat $(ud)$ or $(uu)$ pair as a diquark. If we consider the probability of the two cases, we naively conclude that the
probability of $(ud)$ as a diquark is two times that of $(uu)$ as a diquark. It should be stressed that we also explicitly take the internal quark structure of the
axial-vector diquark into account. This treatment is different from the non-relativistic quark model calculations for the nucleon EMFFs and for
the $N-\Delta$ transition amplitudes, where the total contribution is simply regarded as three times of the single quark contribution although
the bound state wave function is employed~\cite{Lipes:1972nf,Capstick:1992uc,Capstick:1992xn,Giannini:2002tn}. The present approach is consistent with
the other relativistic and covariant quark-diquark approaches~\cite{Meyer:1994cn,Keiner:1995bu}. \\

\subsection{EMFFs of $\Delta$ contributed by quark}\label{sub3_1}
\par\noindent\par
\begin{figure}[htbp]
    \begin{center}
        \subfigure[]{\includegraphics[scale=0.5]{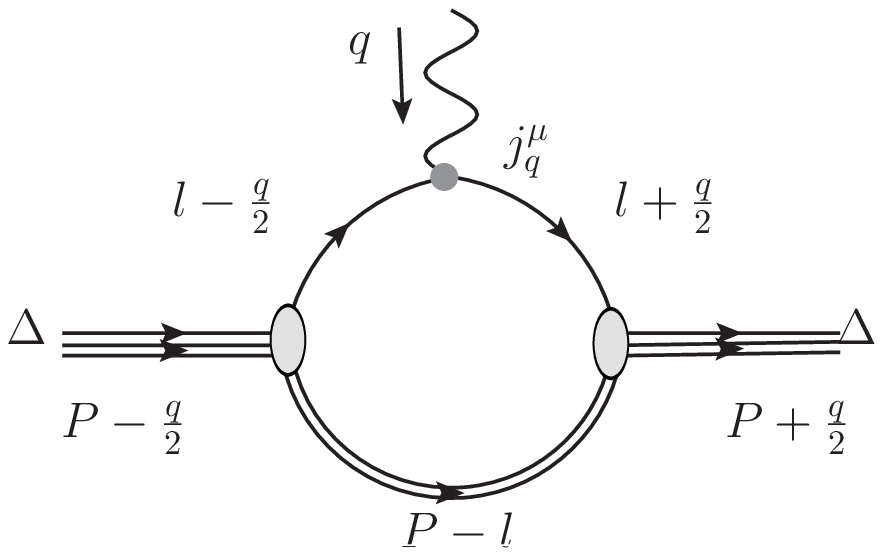}}
        \subfigure[]{\includegraphics[scale=0.5]{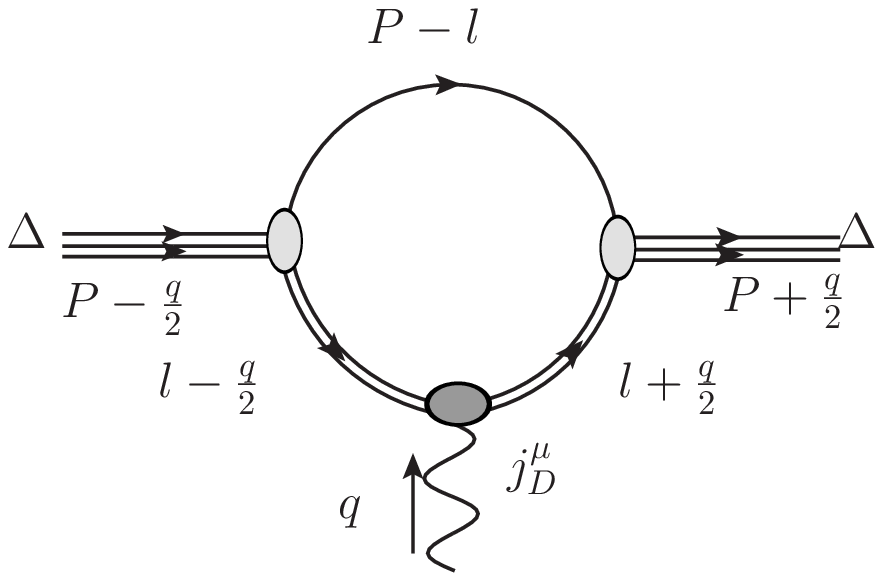}}
        \subfigure[]{\includegraphics[scale=0.5]{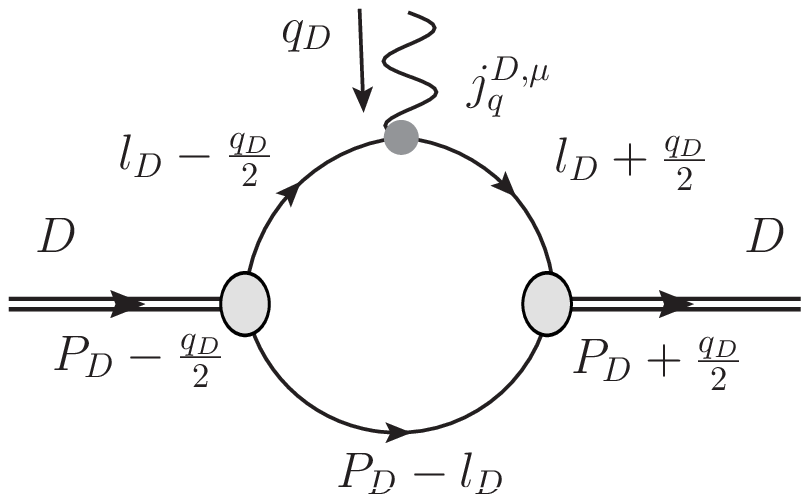}}
        \caption{\small{Feynman diagrams for the electromagnetic current of the $\Delta$ resonance, (a) and (b), and
         of the diquark (c). The left and middle panels stand for the contributions of quark (single line) and diquark (double line) to $\Delta$.}}
        \label{FeynmanDeltaEM}
    \end{center}
\end{figure}

Here, we give the details for the calculation of the EMFFs of $\Delta$ in our approach. The electromagnetic current attached to $\Delta$
is represented by the Feynman diagrams illustrated in Figs.~\ref{FeynmanDeltaEM} (a) and (b) and its matrix element is expressed as the sum
of the quark and diquark contributions (labeled by the subscripts of $q$ and $D$, respectively) as
\begin{equation}
    \left\langle p^\prime,\lambda^\prime \left| \hat{J}^{\mu}(0) \right| p,\lambda\right\rangle = \left\langle p^\prime,\lambda^\prime
    \left| \hat{J}^{\mu}_{q}(0) \right| p,\lambda\right\rangle + \left\langle p^\prime,\lambda^\prime \left| \hat{J}^{\mu}_D(0)
    \right| p,\lambda\right\rangle.
\end{equation}
In the present work, we neglect the longitudinal part $k^\mu k^\nu/m_V^2$ of the vector propagator in order to have finite results~\cite{Dong:2009yp}. So the quark contribution is
\begin{equation}\label{DeltaMt}
    \begin{split}
    &\left\langle p^\prime,\lambda^\prime \left| \hat{J}^{\mu}_{q}(0)
    \right| p,\lambda\right\rangle \\
    =& - Q^e_{q} e \bar{u}_{\alpha'}(p',\lambda') (-i) \int \frac{d^4 l}{(2 \pi)^4}\frac{1}{\widetilde{\mathfrak{D}}}
    \tilde{\Gamma}^{\alpha' \beta'} \left( \slashed{l}+\frac{\slashed{q}}{2}+m_q \right)
    g_{\beta' \beta} \gamma^\mu \left( \slashed{l}-\frac{\slashed{q}}{2}
    +m_q \right) \tilde{\Gamma}^{\beta \alpha} u_{\alpha}(p,\lambda),
    \end{split}
\end{equation}
where $Q^e_{q}$ is the charge quantum number carried by the active quark, and $\widetilde{\mathfrak{D}}$ stands for all the propagator denominators as
\begin{equation}\label{DeltaDt}
    \begin{split}
    \widetilde{\mathfrak{D}}=&\biggl[\left(l+ \frac{q}{2} \right)^2 - m_q^2+i \epsilon\biggr]
    \biggl[\left(l- \frac{q}{2} \right)^2 - m_q^2+i \epsilon\biggr]
    [\left(l-P\right)^2-m_D^2+i \epsilon].
    \end{split}
\end{equation}
The vertex of $\Delta$ with its quark and diquark constituents in Eq.~\eqref{DeltaMt} is expressed as
$ \tilde{\Gamma}^{\alpha \beta} = \Gamma^{\alpha \beta} \Xi$.
According to Ref.~\cite{Scadron:1968zz}, the Lorentz structure of the vertex $\Gamma^{\alpha\beta}$ is
\begin{equation}\label{vertexfunction}
    \Gamma^{\alpha \beta} =c_1\big [ g^{\alpha \beta} + g_2 \gamma^\beta \Lambda^\alpha
    + g_3 \Lambda^\beta \Lambda^\alpha\big ], 
\end{equation}
with $\Lambda$ being the relative momentum between the quark and diquark. The couplings of $c_1$, $g_2$, and $g_3$ in Eq.~\eqref{vertexfunction}
can be determined by fitting to the experimental data of EMFFs or to the Lattice calculation. The superscript $\beta$ stands for the index of
the spin-1 particle. It should be addressed
that the vertex $\Gamma^{\alpha\beta}$ contains high-order momentum terms, and they can make the loop integral divergent. To avoid this problem
we simply consider an additional scalar function $\Xi$ to simulate the bound state problem of the $\Delta$ resonance. In general, this scalar
function should be obtained from a dynamical calculation of the system, like solving the Bethe-Salpeter equation. Here, we simply
take an ansatz for the scalar function $\Xi$ as~\cite{Frederico:2009fk}
\begin{equation}\label{vertexfunction2}
    \Xi (p_1,p_2)=\frac{c}{[ p_1^2 -m_R^2+i \epsilon][ p_2^2 -m_R^2+i \epsilon]},
\end{equation}
where $m_R$ is a cut-off mass parameter, and we find that our numerical results are not sensitive to $m_R$ within a certain range.
Then, Eq.~\eqref{DeltaMt} goes to
\begin{equation}\label{EMFFs_quark}
    \begin{split}
    &\left\langle p^\prime,\lambda^\prime \left| \hat{J}^{\mu}_{q}(0)
    \right| p,\lambda\right\rangle =- Q^e_{q} e \bar{u}_{\alpha'}(p',\lambda') {\left( -i {\tilde C}^2 \right)} \\
    ~~~&\times \int \frac{d^4 l}{(2 \pi)^4}\frac{1}{\mathfrak{D}}
    \Gamma^{\alpha' \beta'} \left( \slashed{l}+\frac{\slashed{q}}{2}+m_q \right)
    g_{\beta' \beta} \gamma^\mu \left( \slashed{l}-\frac{\slashed{q}}{2}
    +m_q \right)\Gamma^{\beta\alpha } u_{\alpha}(p,\lambda),
    \end{split}
\end{equation}
where ${\tilde C}=cc_1$ 
and denominator is modified to be
\begin{equation}\label{DeltaD}
    \begin{split}
    \mathfrak{D}=\widetilde{\mathfrak{D}} [\left( l-P \right) ^2 -m_R^2+i \epsilon]^2
    \biggl[\left( l-\frac{q}{2} \right) ^2 -m_R^2+i \epsilon\biggr]
    \biggl[\left( l+\frac{q}{2} \right) ^2 -m_R^2+i \epsilon\biggr].
    \end{split}
\end{equation}

\subsection{EMFFs of $\Delta$ contributed by diquark}\label{sub3_2}
\par\noindent\par
In the same way, the diquark contribution to the EMFFs of $\Delta$ is
\begin{equation}
    \begin{split}
    &\left\langle p^\prime,\lambda^\prime \left| \hat{J}^{\mu}_D(0)
    \right| p,\lambda\right\rangle \\
    &= - Q^e_{D} e \bar{u}_{\alpha'}(p',\lambda') {i {\tilde C}^2} \int \frac{d^4 l}{(2 \pi)^4}\frac{1}{\mathfrak{D}'}
    \Gamma^{\alpha'}_{~\beta'} \left( \slashed{P}-\slashed{l}+m_q \right)
     j_D^{\mu,\beta' \beta} \Gamma^{~\alpha}_{\beta} u_{\alpha}(p,\lambda),
    \end{split}
\end{equation}
where the $Q^e_{D}$ is the charge quantum number carried by the diquark. It should be mentioned that the diquark is an axial vector $(1^+)$ bound state
of two quarks. Here, we adopt the same vertex in Eq.~\eqref{vertexfunction} and Eq.~\eqref{vertexfunction2}. Thus, in the above equation,
\begin{equation}\label{DeltaD'}
    \begin{split}
    \mathfrak{D}'=&\biggl[\left(l+ \frac{q}{2} \right)^2 - m_D^2+i \epsilon\biggr]
    \biggl[\left(l- \frac{q}{2} \right)^2 - m_D^2+i \epsilon\biggr]
    [\left(l-P\right)^2-m_q^2+i \epsilon][\left( l-P \right) ^2 -m_R^2+i \epsilon]^2\\
    &\biggl[\left( l-\frac{q}{2} \right) ^2 -m_R^2+i \epsilon\biggr]
    \biggl[\left( l+\frac{q}{2} \right) ^2 -m_R^2+i \epsilon\biggr].
    \end{split}
\end{equation}
\\
Figure~\ref{FeynmanDeltaEM}~(c) gives the explicit contribution of diquark with its quark structure. The effective Lagrangian for the diquark
is~\cite{Meyer:1994cn}
\eq
{\cal L}_{D\to qq}=c_D\Psi_q^TC^{-1}\gamma^{\mu}\Psi_q\epsilon_{\mu,D}(p_D,\lambda)\Xi_D~+~H.~C.,
\en
where $\Psi_q^T$ stands for the charge conjugate of quark field and $C=i\gamma^2\gamma^0$.
The correlation function attached to the vertex in the above Lagrangian and Fig.~\ref{FeynmanDeltaEM}~(c) is assumed to be the same as in
Eq.~\eqref{vertexfunction2} for simplicity and with the same cut-off mass $m_R$ in order to reduce the number of parameters.
According to Fig.~\ref{FeynmanDeltaEM}~(c), we get
\begin{equation}\label{ElmcD}
    \begin{split}
        \left\langle p^\prime,\lambda^\prime \left| \hat{J}_D^{\mu}(0) \right| p,\lambda\right\rangle = \Sigma_{q} \left\langle p^\prime,\lambda^\prime
        \left| \hat{J}^{\mu}_{q}(0) \right| p,\lambda\right\rangle
        = - \epsilon^*_{\beta ^\prime}\left(p_D^\prime,\lambda^\prime\right) j_D^{\mu, \beta' \beta} \epsilon_\beta \left( p_D,\lambda\right),
    \end{split}
\end{equation}
where $\epsilon_\beta \left( p_D,\lambda\right)$ represents the spin-1 diquark field, and $j_D^{\mu, \beta \beta'}$ represents the effective electromagnetic
current of spin-1 diquark. Here, we introduce kinematical variables $P_D^\mu = (p_D^\mu + p_D'^\mu)/2, q_D^\mu = p_D'^\mu-p_D^\mu = q^\mu$
and $q_D^2=-t_D=-t$ (since transfer momentum is all on diquark). Then
\begin{equation}\label{DiquarkEMFFs}
    \begin{split}
    &\left\langle p_D^\prime,\lambda^\prime \left| \hat{J}^{\mu}_{q}(0)
    \right| p_D,\lambda\right\rangle  = - Q^e_{q} e \epsilon^*_{\beta'}(p_D',\lambda') i {\tilde C}_D^2\\
    &\times  \int \frac{d^4 l_D}{(2 \pi)^4}\frac{1}{\mathfrak{D}_D}
    \gamma^{\beta'} \left( \slashed{l}_D+\frac{\slashed{q}}{2}+m_q \right)
     \gamma^\mu \left( \slashed{l}_D-\frac{\slashed{q}}{2}
    +m_q \right)\gamma^{\beta} (\slashed{l}_D - \slashed{P}_D + m_q) \epsilon_{\beta}(p_D,\lambda),
    \end{split}
\end{equation}
where the constant ${\tilde C}$ in Eq.~\eqref{EMFFs_quark} is replaced by ${\tilde C}_D=c_1c_D$. In addition,
\begin{equation}\label{DeltaDD}
    \begin{split}
    \mathfrak{D}_D=&\biggl[\left(l_D+ \frac{q}{2} \right)^2 - m_q^2+i \epsilon\biggr]
    \biggl[\left(l_D- \frac{q}{2} \right)^2 - m_q^2+i \epsilon\biggr]
    [\left(l_D-P_D\right)^2-m_q^2+i \epsilon][\left( l_D-P_D \right) ^2 -m_R^2+i \epsilon]^2\\
    &\biggl[\left( l_D-\frac{q}{2} \right) ^2 -m_R^2+i \epsilon\biggr]
    \biggl[\left( l_D+\frac{q}{2} \right) ^2 -m_R^2+i \epsilon\biggr].
    \end{split}
\end{equation}
Finally, the electromagnetic current of the diquark in Eq.~\eqref{ElmcD} can be written as
\begin{equation}\label{Diquark_EMFFs}
    j_D^{\mu , \beta' \beta} = \Big [ g^{\beta' \beta} F_{D; 1}^V(t) - \frac{q^{\beta'} q^{\beta}}{2 m_D^2} F_{D; 2}^V(t) \Big ] (p'_D+p_D)^\mu
    - (q^{\beta'} g^{\mu \beta}   - q^{\beta} g^{\mu \beta'}) F_{D; 3}^V(t),
\end{equation}
\noindent
where $F_{D; 1,2,3}^V(t)$ stand for the three form factors of the spin one particle contributed by quarks and by the loop integral. They contain the
binding effect. The expression of this effective current in Eq.~\eqref{Diquark_EMFFs} is standard for a free spin-1 particle. Moreover, in reproducing the effective EM
current, the normalization of the diquark charge is also employed. \\

\subsection{GFFs of the $\Delta$ contributed by the quark}\label{sub3_3}
\par\noindent\par
\begin{figure}[ht]
    \begin{center}
        \subfigure[]{\includegraphics[scale=0.5]{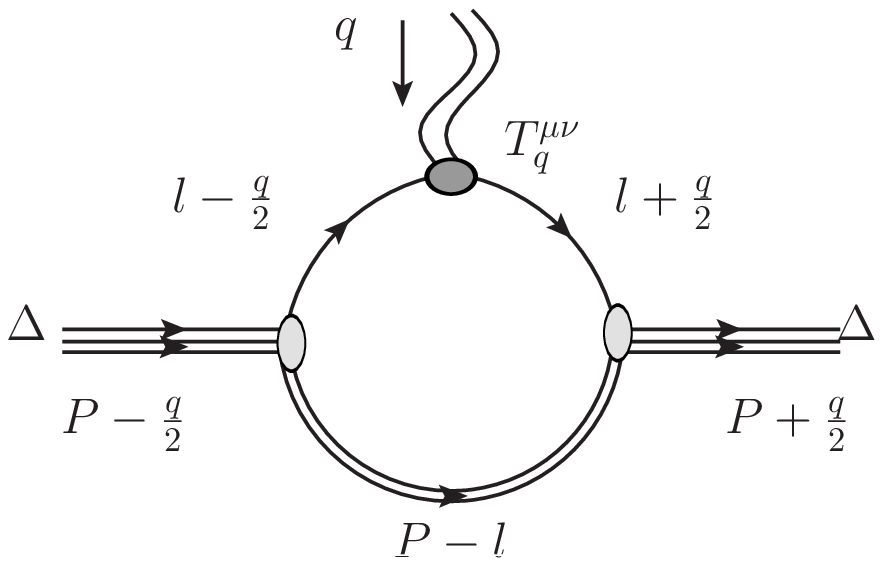}}
        \subfigure[]{\includegraphics[scale=0.5]{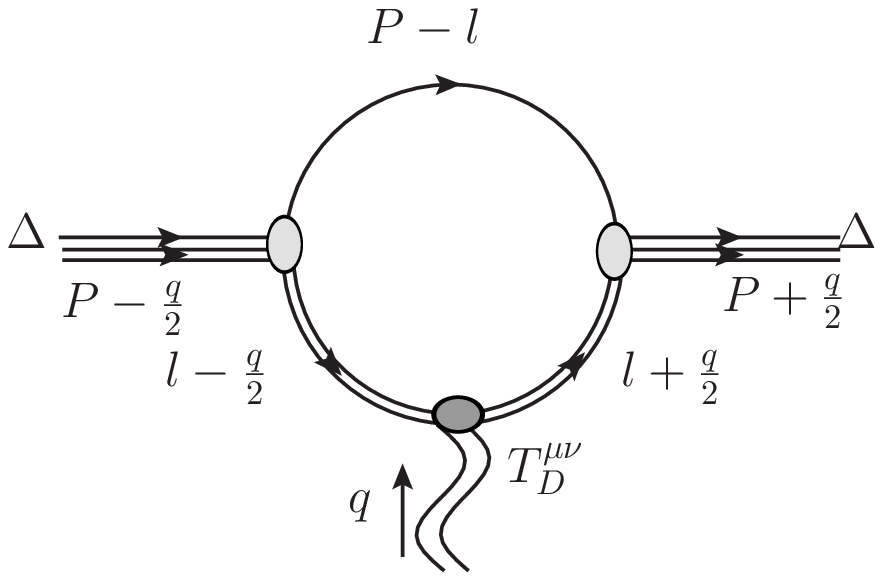}}
        \subfigure[]{\includegraphics[scale=0.5]{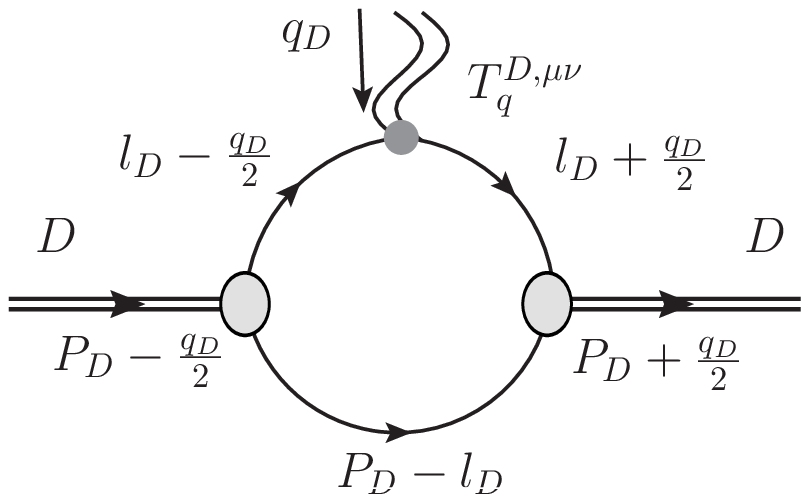}}
        \caption{\small{Feynman diagrams for the GFFs of $\Delta$, contributed by quark (a) and by diquark (b), and
         the GFFs of the diquark (c).}}
       \label{FeynmanDeltaEMT}
    \end{center}
\end{figure}

One may also calculate the matrix elements of energy-momentum tensor for the $\Delta$ system by summing the contributions of the quark and the diquark:
\begin{equation}
    T^{\mu \nu} = T^{\mu \nu}_q + T^{\mu \nu}_D.
\end{equation}
The Feynman diagrams for the process are shown in Figs.~\ref{FeynmanDeltaEMT} (a) and (b).

According to the Lagrangian for a quark with mass $m_q$
\begin{equation}
    \mathcal{L}_{q} = \frac{i}{2} \bar{\psi}_q \gamma^\mu \overleftrightarrow{\partial}_\mu \psi_q
    -m_q \bar{\psi}_q \psi_q, \quad \mbox{with} \quad \overleftrightarrow{\partial}_\mu =
    \overrightarrow{\partial}_\mu - \overleftarrow{\partial}_\mu,
\end{equation}
then, the symmetric EMT is defined as
\begin{equation}\label{Tquark}
    T_q^{\mu \nu}=\frac{i}{4} \bar{\psi}_q \gamma^\mu \overleftrightarrow{\partial^\nu} \psi_q + \frac{i}{4} \bar{\psi}_q \gamma^\nu
    \overleftrightarrow{\partial^\mu} \psi_q .
\end{equation}
\par\noindent\par
In our covariant quark-diquark approach, the matrix element of EMT current from the contribution of quark is
\begin{equation}\label{Dterm}
    \begin{split}
    &\left\langle p^\prime,\lambda^\prime \left| \hat{T}^{\mu \nu}_{q}(0)
    \right| p,\lambda\right\rangle = - \bar{u}_{\alpha'}(p',\lambda') \frac{-i {\tilde C}^2}{2} \\
    &\times \int \frac{d^4 l}{(2 \pi)^4}\frac{1}{\mathfrak{D}}
    \Gamma^{\alpha'\beta'} \left( \slashed{l}+\frac{\slashed{q}}{2}+m_q \right)
    g_{\beta' \beta} (\gamma^\mu l^\nu + \gamma^\nu l^\mu ) \left( \slashed{l}-\frac{\slashed{q}}{2}
    +m_q \right)\Gamma^{\alpha\beta} u_{\alpha}(p,\lambda),
    \end{split}
\end{equation}
where $\mathfrak{D}$ has been given in Eq.~\eqref{DeltaD}.

\subsection{GFFs of the $\Delta$ contributed by the diquark}\label{sub3_4}
\par\noindent\par
The EMT of the diquark can be obtained from the Proca Lagrangian if we consider it as a structureless particle.
Here, we treat the diquark contribution to the EMT matrix element of $\Delta$ by considering explicitly its quark contents as we have discussed
in Sec. 3.2. The matrix element of EMT current from the contribution of diquark is
\begin{equation}\label{DDterm}
    \begin{split}
    &\left\langle p^\prime,\lambda^\prime \left| \hat{T}^{\mu \nu}_{D}(0)
    \right| p,\lambda\right\rangle \\
    =& - \bar{u}_{\alpha'}(p',\lambda') i {\tilde C}^2 \int \frac{d^4 l}{(2 \pi)^4}\frac{1}{\mathfrak{D}'}
    \Gamma^{\alpha' \beta'} \left( \slashed{l}+\frac{\slashed{q}}{2}+m_q \right)
     X_{\beta' \beta}^{~~~\mu \nu} \left( \slashed{l}-\frac{\slashed{q}}{2}
    +m_q \right)\Gamma^{\beta \alpha } u_{\alpha}(p,\lambda),
    \end{split}
\end{equation}
where $\mathfrak{D}'$ is shown in Eq.~\eqref{DeltaD'}, and $X_{\beta'\beta}^{\mu\nu}$ stands for the effective energy-momentum tensor of the diquark. \\

According to Fig.~\ref{FeynmanDeltaEMT}~(c), the matrix element of the EMT current of the diquark, due to its two quark structure,
is~\cite{Cosyn:2019aio,Polyakov:2019lbq}
\begin{equation}
    \begin{split}
    &\left\langle p_D^\prime,\lambda^\prime \left| \hat{T}^{\mu \nu}_D(0)
    \right| p_D,\lambda\right\rangle
    = 2 \left\langle p_D^\prime,\lambda^\prime \left| \hat{T}^{\mu \nu}_q(0)
    \right| p_D,\lambda\right\rangle
    =  - \epsilon^*_{\beta ^\prime}\left(p_D^\prime,\lambda^\prime\right) X^{\beta' \beta \mu \nu} \epsilon_\beta \left( p_D,\lambda\right)\\
    &~~~= - \epsilon^*_{\beta ^\prime}\left(p_D^\prime,\lambda^\prime\right) \bigg [\frac{ P_D^\mu P_D^\nu}{m_D} \left(g^{\beta  \beta '} F_{D; 1,0}^{T} (t)
    -\frac{q ^{\beta } q ^{\beta'}}{2 m_D^2}F_{D; 1,1}^{T} (t) \right) \\
    &~~~~~~+ \frac{ \left({q }^\mu {q }^\nu- {g}^{\mu  \nu } q ^2\right)}{4m_D}
    \left({g}^{\beta  \beta '}F_{D; 2,0}^{T} (t)-\frac{{q }^{\beta }
    {q }^{\beta '}}{2 m_D^2}F_{D; 2,1}^{T} (t)\right) \\
    &~~~~~~
    - \frac{ P_D^{\{ \mu}g^{\nu \} [ \beta'} q^{\beta ]}}{m_D} F_{D; 4,0}^{T} (t) 
    - \frac{1}{m_D} \left({q }^{\{ \mu} g^{\nu \} \{ \beta '} {q }^{\beta \}}
    -2 q^{\beta } q^{\beta '} g^{\mu  \nu }
    - g^{\beta ' \{ \mu } g^{\nu \} \beta } q^2 \right) F_{D; 5,0}^{T} (t)
    \bigg ]
    \epsilon_\beta \left( p_D,\lambda\right),
    \end{split}
\end{equation}
where the nonconserving form factors are ignored. Finally, we get the matrix element of EMT tensor contributed by the diquark
\begin{equation}\label{DiquarkGFFs}
    \begin{split}
    &\left\langle p_D^\prime,\lambda^\prime \left| \hat{T}^{\mu \nu}_{D}(0)
    \right| p_D,\lambda\right\rangle = - \epsilon^*_{\beta'}(p_D',\lambda') i \tilde C_D^2 \\
    &\times \int \frac{d^4 l_D}{(2 \pi)^4}\frac{1}{\mathfrak{D}_D}
    \gamma^{\beta'} \left( \slashed{l}_D+\frac{\slashed{q}}{2}+m_q \right)
     (\gamma^\mu l_D^\nu + \gamma^\nu l_D^\mu) \left( \slashed{l}_D-\frac{\slashed{q}}{2}
    +m_q \right)\gamma^{\beta} (\slashed{l}_D - \slashed{P}_D + m_q) \epsilon_{\beta}(p_D,\lambda).
    \end{split}
\end{equation}
\vspace{0.5cm}
\par\noindent\par
To summarize this section, we employ the relativistic covariant quark-diquark approach to compute the EMFFs and GFFs
of the spin-3/2 $\Delta$ resonance. In the above formulas the quark structure of the diquark ($1^+$)
is explicitly taken into account by introducing the correlation function and by the loop integrals of
Eqs.~\eqref{DiquarkEMFFs} and~\eqref{DiquarkGFFs}. In particular, only the fundamental electromagnetic current and the EMT of the quark are
involved. \\

\section{Numerical results}\label{section4}
\subsection{Determination of model parameters}
\par\noindent\par
In the present approach, we need to numerically calculate the loop integrals sandwiched between the two Rarita-Schwinger spinors. The on-shell identities
which have been explicitly proven in Ref.~\cite{Cotogno:2019vjb} for the Rarita-Schwinger spinors are employed. They are listed in Appendix~A.
Moreover, Appendix~B gives the Feynman parameterizations for the necessary loop integrals.\\

We also need to input the masses of the $\Delta$ resonance $M$, quark $m_q$, diquark $m_D$, and the cut-off $m_R$ in the calculation. Here, we simple
choose $M=1.085~\text{GeV}$. It is the average of the masses of nucleon and $\Delta$ resonance, and this selection means that we do not
consider the mass-splitting between the $\Delta$ and nucleon. Moreover, we assume $m_q=0.4~\text{GeV}$ according to Ref.~\cite{Cloet:2014rja}.
Our $M$ and $m_q$ indicate that $M~<~3m_q$. Furthermore, we choose $m_D\sim 0.76~\text{GeV}$~\cite{Cloet:2014rja}, it implies that the diquark
is a bound state of two quarks as well. Finally we simply borrow $m_R\sim ~1.6~\text{GeV}$ from Ref.~\cite{Sun:2017gtz}. \\

\begin{figure}[ht]
    \begin{center}
        \includegraphics[height=4.5cm]{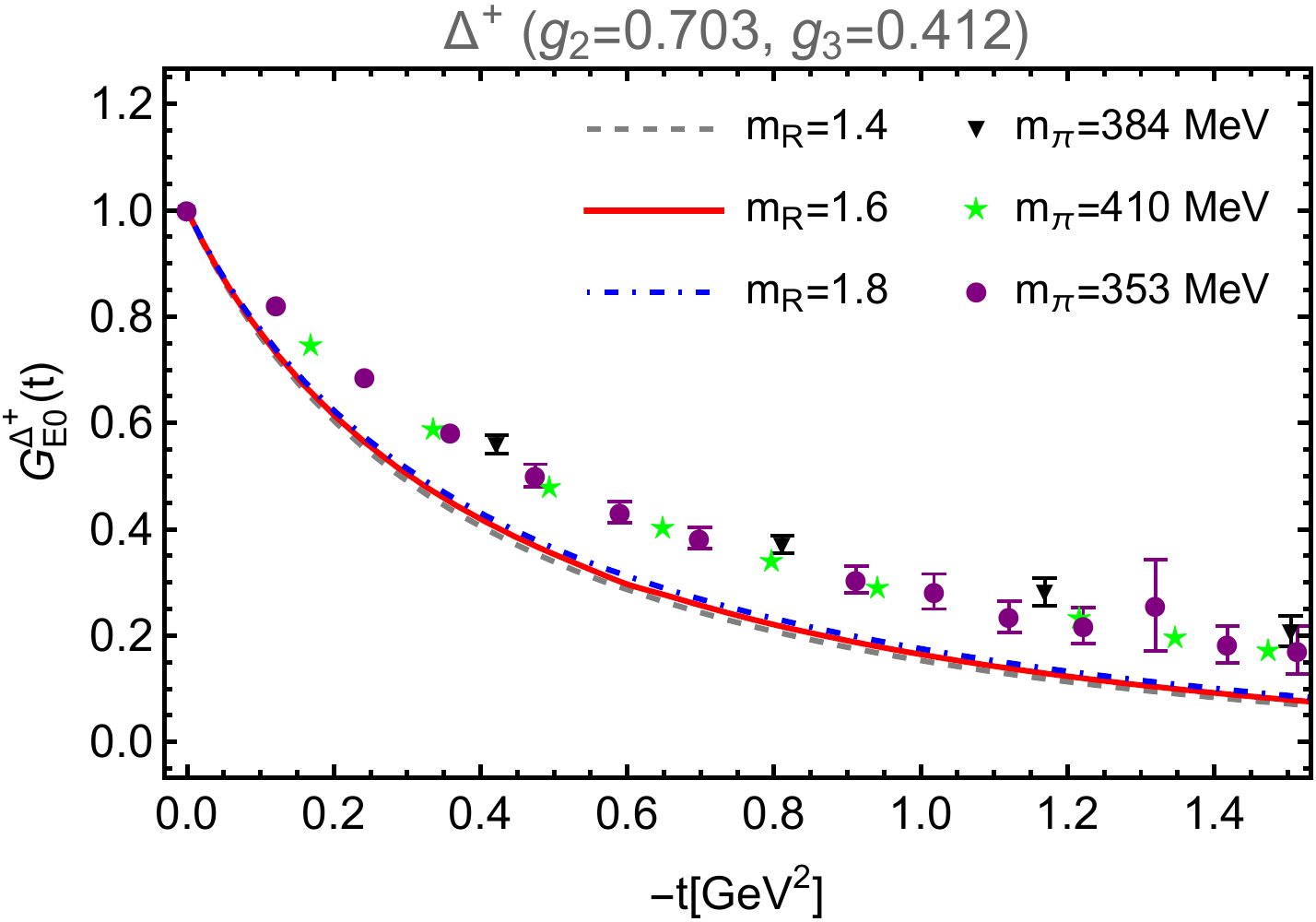}
        \includegraphics[height=4.5cm]{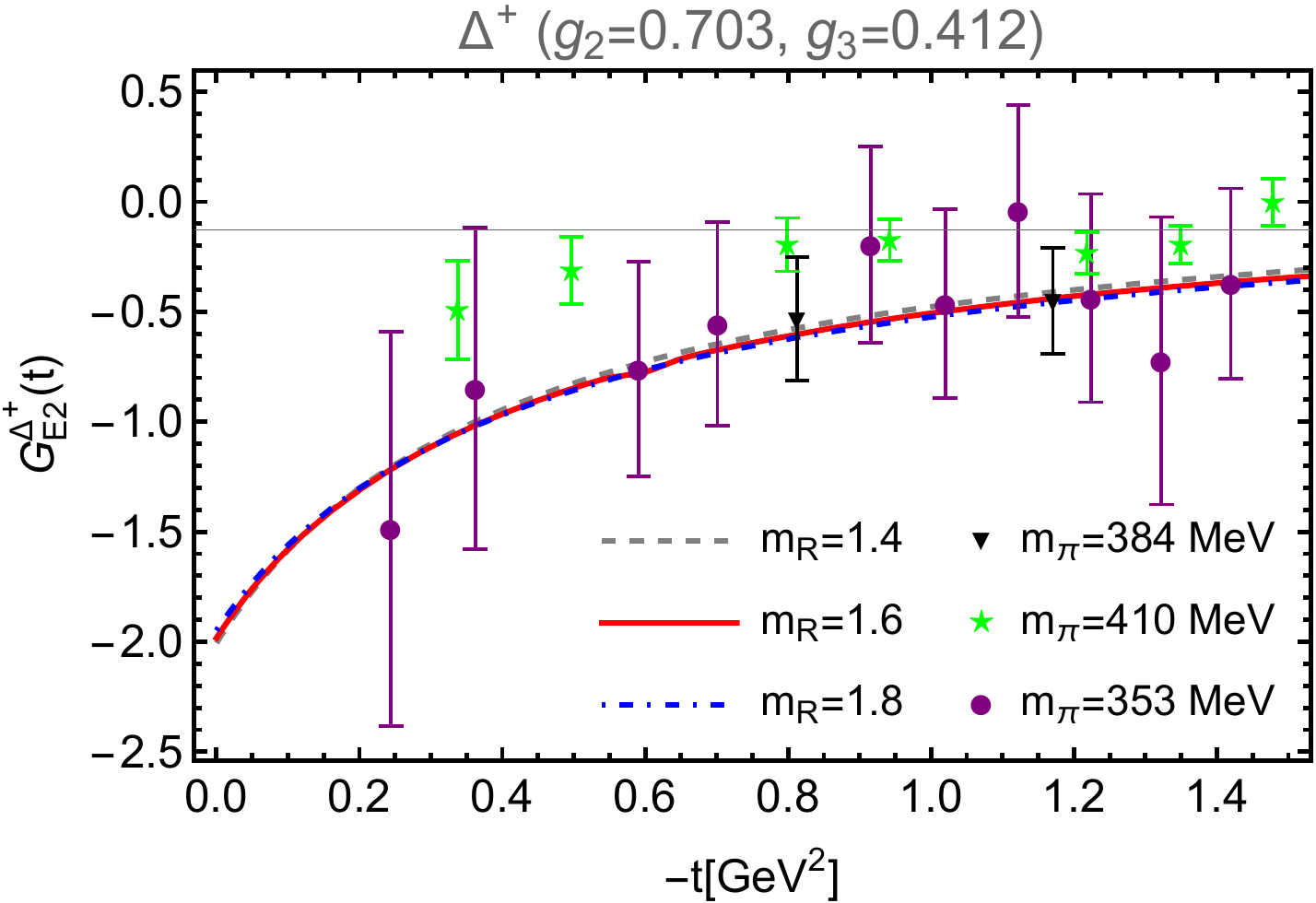}
        \includegraphics[height=4.5cm]{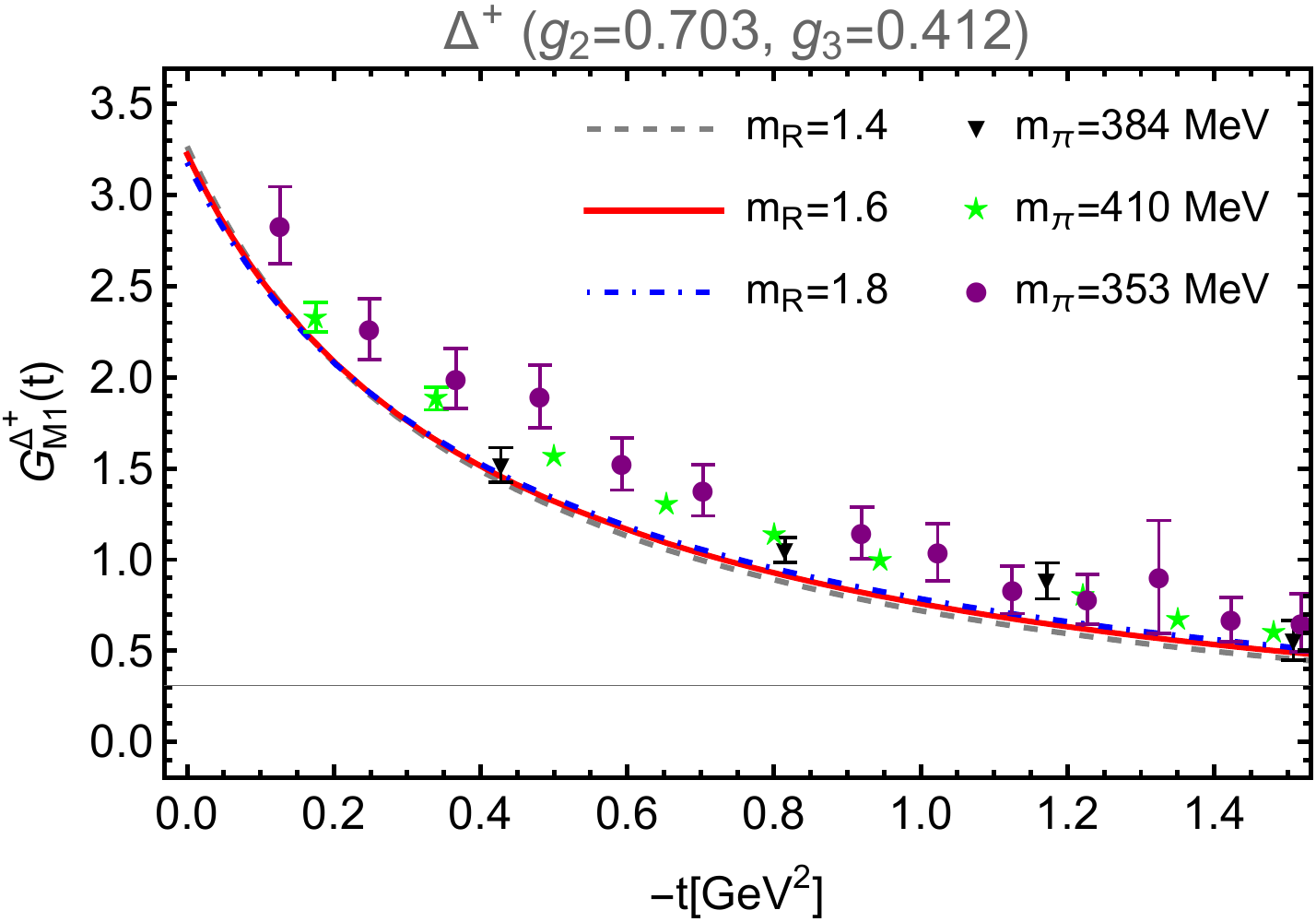}
        \includegraphics[height=4.5cm]{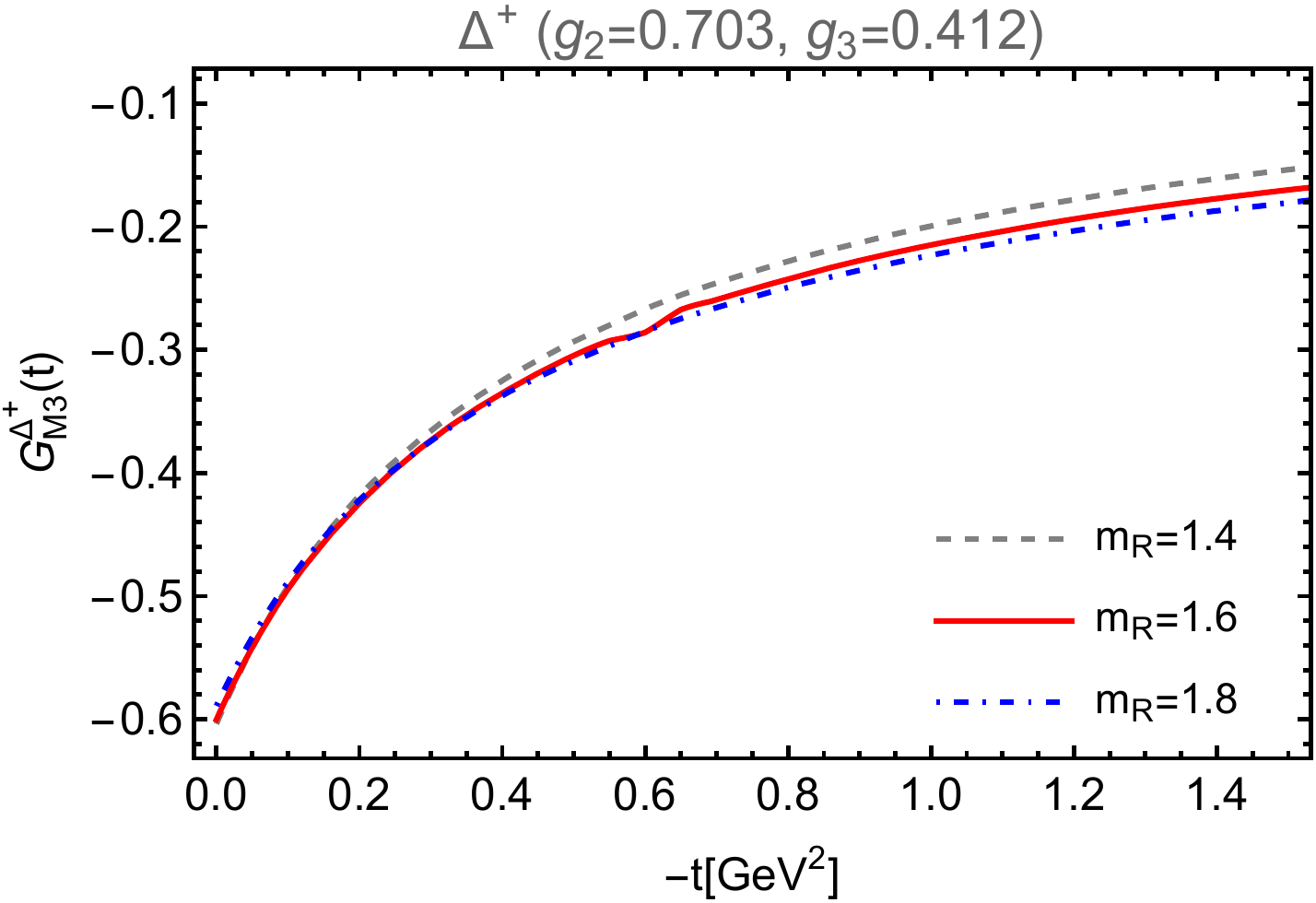}
        \caption{\small{Calculated four EMFFs of $\Delta^+$ comparing to the Lattice QCD calculations. The dashed, solid and the dotted-dashed
         curves represent the results with $m_R=1.4, 1.6$ and $1.8~\text{GeV}$, respectively, and $g_2=0.412~\text{GeV}^{-1}$ and
         $g_3=0.703~\text{GeV}^{-2}$ are used.}}
        \label{g2g3}
    \end{center}
\end{figure}

It should be addressed that due to the normalization of the charge form factor of $\Delta$ at $t^2=0$, the overall factor ${\tilde C}=cc_1$ can be
fixed. However, $g_2$ and $g_3$ in Eq.~\eqref{EMFFs_quark} are still free. They describe the $D$-wave coupling of the $\Delta$ resonance to the quark
and the axial vector diquark in our approach, and they provide an essential effect on the high-order multipoles. To determine these two parameters,
the EMFFs calculated by the Lattice QCD (LQCD) of Ref.~\cite{Alexandrou:2008bn} are employed as constraints. Comparing to the LQCD results, we select
$g_2 = 0.703~\text{GeV}^{-1}$ and $g_3 = 0.412~\text{GeV}^{-2}$.
All the parameters in our calculation are listed in Tab.~\ref{parameterstable}, and the obtained
four-EMFFs are plotted in Fig.~\ref{g2g3} for $\Delta^+$. Figure~\ref{g2g3} shows that our results are consistent with the LQCD calculation,
at least qualitatively. In the figure, the lines are our calculations with different cut-off masses and the dots are the results from LQCD
with different pion masses. We also conclude that our results are not sensitive to the cut-off parameter $m_R$.
In the present work, the units of parameters in figures have been omitted and are consistent with Tab.~\ref{parameterstable}.\\

\begin{table}[ht]
    \centering
	\begin{tabular}{ | c | c | c | c | c | c | c |}
		\hline
		$M/\text{GeV}$	& $m_q/\text{GeV}$	& $m_D/\text{GeV}$	& $m_R/\text{GeV}$	& $g_2/\text{GeV}^{-1}$	& $g_3/\text{GeV}^{-2}$\\
		\hline
		1.085	& 0.4	& 0.76	& 1.6	& 0.703	& 0.412	\\
		\hline
	\end{tabular}
    \caption{\small{The parameters used in our approach.}}
    \label{parameterstable}
\end{table}

\begin{figure}[hbp]
    \begin{center}
        \subfigure[]{\includegraphics[height=4.5cm]{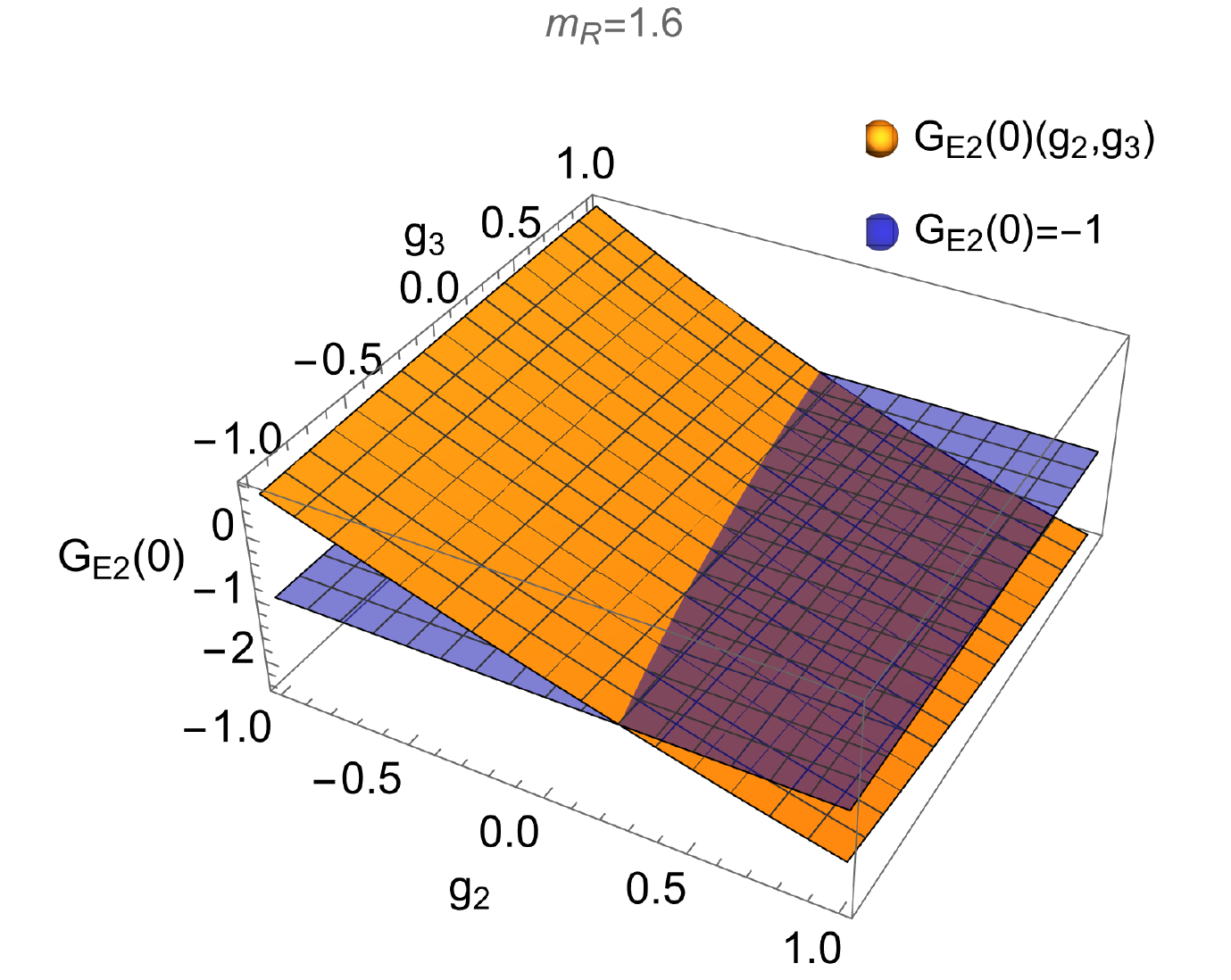}}
        \subfigure[]{\includegraphics[height=4.5cm]{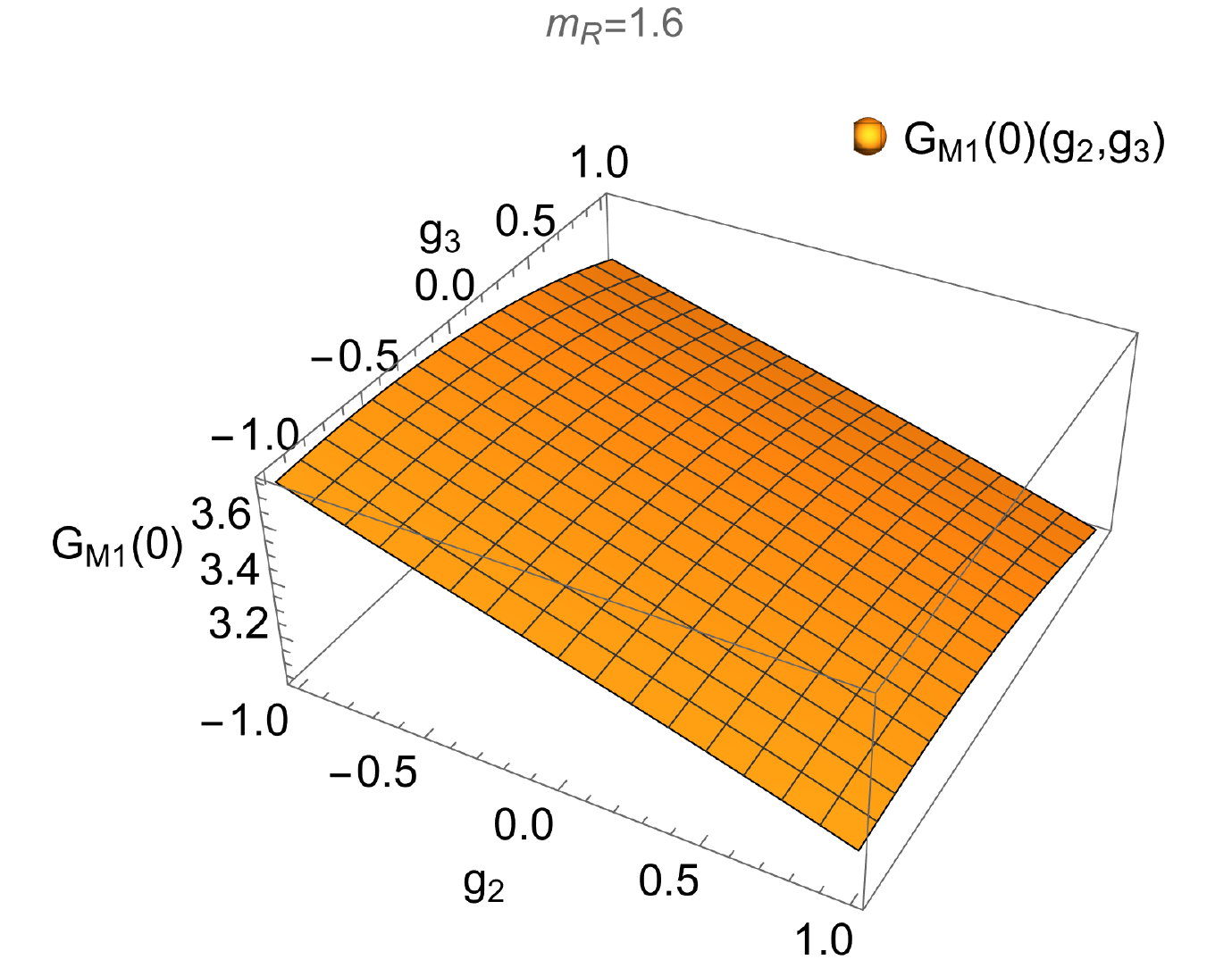}}
        \subfigure[]{\includegraphics[height=4.5cm]{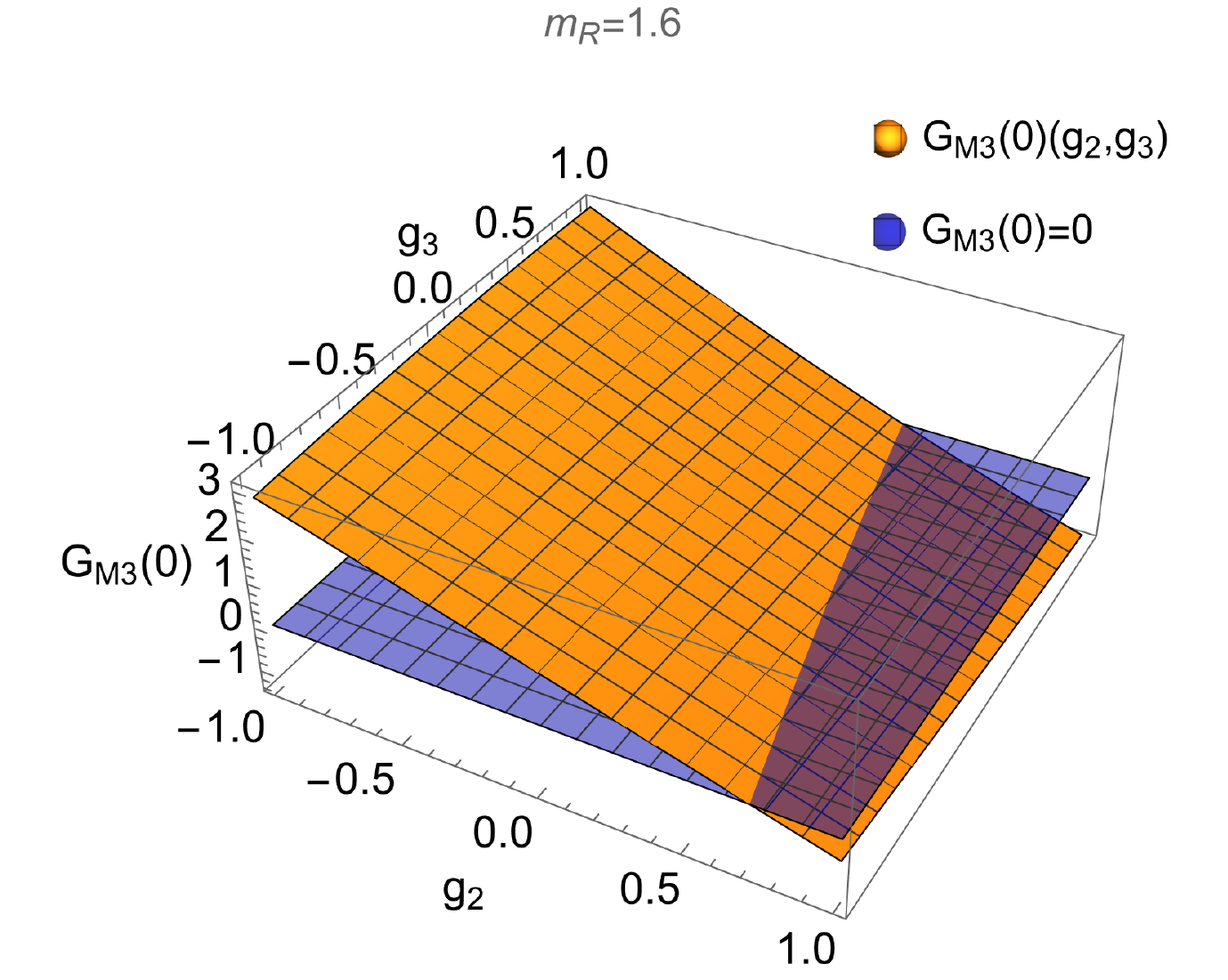}}
        \caption{\small{The parameter $g_2$ and $g_3$ dependences of $G_{E2}(0)$ (a), $G_{M1}(0)$ (b), and $G_{M3}(0)$ (c).}}
        \label{mR}
    \end{center}
\end{figure}
To show a more detailed analysis of our model parameters, we, first of all, check the impact of $g_2$ and $g_3$ on our EMFFs.
Figure~\ref{mR} displays their effect. We find that $g_2$ and $g_3$ have a remarkable influence to the electric-quadrupole and magnetic-octupole
form factors, and they even change the signs of $G_{E2}$ and $G_{M3}$. However, they have a little impact on the electric monopole and
magnetic-dipole form factors. This conclusion is reasonable since the couplings $g_{2,3}$ stand for the high-partial waves, and they manifest themselves
in the high-order multipoles, like quadrupole and octupole form factors. \\

\subsection{Results for the moments of EMFFs}
\par\noindent\par
When the squared momentum transfer goes to zero $t=0$, the form factors give the moments of the magnetic-dipole
$\mu_{\Delta} = G_{M1}(0)\frac{e}{2M}$, of the electric-quadrupole $\mathcal{Q}_{\Delta} = G_{E2}(0) \frac{e}{M^2}$, and of the magnetic-octupole
$\mathcal{O}_{\Delta} = G_{M3}(0) \frac{e}{2M^3}$, where $e$ is the electric charge~\cite{Ramalho:2009vc}. We can compare the obtained
magnetic-dipole, electric-quadrupole and magnetic-octupole moments to the results of different model calculations, such as non-relativistic
quark model (NQM)~\cite{ParticleDataGroup:1992tph,Isgur:1981yz,Krivoruchenko:1991pm},
relativistic quark model (RQM)~\cite{PhysRevD.48.4478},
QCD sum rules (QCDSR)~\cite{Lee:1997jk,Zhu:1998aj,Dey:1999fi,Azizi:2009egn},
light cone QCD sum rules (LCQSR)~\cite{Aliev:2000cy},
Large $N_c$~\cite{Jenkins:1994md,Luty:1994ub,Buchmann:2002et},
chiral quark model with meson exchange currents ($\chi$QMEC)~\cite{Wagner:2000ii,Buchmann:2004ia},
QCD quark model (QCDQM)~\cite{PhysRevD.58.113003,an:2006zf},
chiral bag model (CBM)~\cite{Hong:2007pr},
general parameterization QCD (GPQCD)~\cite{Buchmann:2008zza},
chiral quark-soliton model ($\chi$QSM)~\cite{Ledwig:2008es},
effective mass and screened charge scheme (EMS)~\cite{Bains:2002cp,Dhir:2009ax},
chiral perturbation theory ($\chi$PT)~\cite{Flores-Mendieta:2009vss,Geng:2009ys,Butler:1993ej},
lattice QCD (LQCD)~\cite{Boinepalli:2009sq,Aubin:2008qp,CSSM:2014knt},
and chiral constituent quark model ($\chi$CQM)~\cite{Girdhar:2015gsa}.
Tables~\ref{Magneticdipoleresults},~\ref{Electroquadrupoleresults}, and~\ref{Magneticoctupoleresults} list the comparisons of our magnetic,
quadrupole, and octupole moments with other model calculations, respectively.\\
\linespread{1.}
\begin{table}
    \centering
    \begin{tabular}{ | c | c | c | c | c |}
        \hline\hline
        $G_{M1}(0)$    &$\Delta^{++}$  &$\Delta^+$ &$\Delta^0$ &$\Delta^-$\\
        \hline\hline
        This work   &6.04   &3.02  &0.00  &$-3.02$\\
        \hline
        NQM\cite{ParticleDataGroup:1992tph}  &5.56  &2.73    &$-0.09$  &$-2.92$\\
        RQM\cite{PhysRevD.48.4478}  &4.76	& 2.38  &0.00   &$-2.38$\\
        QCDSR\cite{Lee:1997jk,Zhu:1998aj,Dey:1999fi}    &4.39$\pm$1.00    &2.19$\pm$0.50    &0.00   &$-2.19$$\pm$0.50\\
        LCQSR\cite{Aliev:2000cy}    &4.4$\pm0.8$   &2.2$\pm$0.4 &0.0   &$-2.2$$\pm$0.4\\
        Large $N_c$\cite{Jenkins:1994md,Luty:1994ub,Buchmann:2002et}    &5.9(4)   &2.9(2)   &--  &$-2.9(2)$\\
        $\chi$QMEC\cite{Wagner:2000ii,Buchmann:2004ia}  &6.93 &3.47 &0.00   &$-3.47$\\
        QCDQM\cite{PhysRevD.58.113003,an:2006zf}    &5.689	& 2.778 &$-0.134$ &$-3.045$\\
        CBM\cite{Hong:2007pr}   &4.52   &2.12   &$-0.29$  &$-2.69$\\
        EMS\cite{Bains:2002cp,Dhir:2009ax}  &4.56  &2.28    &0  &$-2.28$\\
        $\chi $PT\cite{Flores-Mendieta:2009vss,Geng:2009ys} &5.390   &2.383 &$-0.625$ &$-3.632$\\
        LQCD\cite{Boinepalli:2009sq,Aubin:2008qp,CSSM:2014knt}  &4.91$\pm$0.61  &2.46$\pm$0.31  &0.00   &$-2.46$$\pm$0.31\\
        $\chi$CQM\cite{Girdhar:2015gsa} &5.82$\pm$0.08   &2.63$\pm$0.06 &$-0.56$$\pm$0.09 &$-3.75$$\pm$0.08\\
        \hline\hline
    \end{tabular}
    \caption{\small{A comparison of our magnetic-dipole moment with other models.}}
    \label{Magneticdipoleresults}
\end{table}
\linespread{1.}
\begin{table}\centering
    \begin{tabular}{ | c | c | c | c | c |}
        \hline\hline
        $G_{E2}(0)$    &$\Delta^{++}$  &$\Delta^+$ &$\Delta^0$ &$\Delta^-$\\
        \hline\hline
        This work   &$-3.86$   &$-1.93$  &0.00  &1.93\\
        \hline
        NQM\cite{Isgur:1981yz} &$-3.82$  &$-1.91$  &0  &1.91\\
        NQM\cite{Krivoruchenko:1991pm} &$-3.63$  &$-1.79$  &0  &1.79\\
        $\chi$PT\cite{Butler:1993ej}    &$-3.12$$\pm$1.95 &$-1.17$$\pm$0.78 &0.47$\pm$0.20  &2.34$\pm$1.17\\
        $\chi$QSM\cite{Ledwig:2008es}   &   &$-2.15$  &   & \\
        QCDSR\cite{Azizi:2009egn}   &$-0.0452$$\pm$0.0113 &$-0.0226$$\pm$0.0057 &0  &0.0226$\pm$0.0057\\
        \hline\hline
    \end{tabular}
    \caption{\small{A comparison of our electric-quadrupole moment with other models.}}
    \label{Electroquadrupoleresults}
\end{table}
\linespread{1.}
\begin{table}\centering
    \begin{tabular}{ | c | c | c | c | c |}
        \hline\hline
        $G_{M3}(0)$    &$\Delta^{++}$  &$\Delta^+$ &$\Delta^0$ &$\Delta^-$\\
        \hline\hline
        This work   &$-1.12$   &$-0.56$  &0.00  &0.56\\
        \hline
        GPQCD\cite{Buchmann:2008zza}   &$-11.68$ &$-5.84$  &0  &5.84\\
        QCDSR\cite{Azizi:2009egn}   &$-0.0925$$\pm$0.0234 &$-0.0462$$\pm$0.0117 &0  &0.0462$\pm$0.0117\\
        \hline\hline
    \end{tabular}
    \caption{\small{A comparison of our magnetic-octupole moment with other model calculations.}}
    \label{Magneticoctupoleresults}
\end{table}\\

For $G_{M1}(0)$ of $\Delta^{++}$, the results of other model calculations are in the range of $[4.4 \sim 6.93]$,
and the minimum value $4.4 \pm 0.8$ predicted by the LCQSR and the maximum value 6.93 by the $\chi$QMEC as shown in Tab.~\ref{Magneticdipoleresults}.
Our result 6.04 is much closer to the one given by the Large $N_c$~\cite{Jenkins:1994md,Luty:1994ub,Buchmann:2002mm,Buchmann:2002et}.
For $G_{E2}(0)$ of $\Delta^{++}$ displayed in Tab.~\ref{Electroquadrupoleresults}, there are the minimum value $-3.82$ in the NQM, and the maximum
value $-0.0452 \pm 0.0113$ in the QCDSR. Our result $-3.86$ is slightly smaller than the results given by other models. The negative sign for
$G_{E2}(0)$ is consistent with most of model calculations and indicates that $\Delta$ is oblate deformed. For $G_{M3}(0)$ listed in
Tab.~\ref{Magneticoctupoleresults}, we see that the results from the two different models vary widely and our result is $-1.12$ for $\Delta^{++}$.
Future measurements for the $\Delta$ isobar deformation are expected to discriminates different approaches. From these three qualitative comparisons
we conclude that our results are comparable to most of the models. In addition, in our numerical calculations, we do not consider the small mass
difference between the $u$ and $d$ quarks, and the different moments for the isospin partners of $\Delta$, displayed in
Tabs.~\ref{Magneticdipoleresults},~\ref{Electroquadrupoleresults}, and~\ref{Magneticoctupoleresults}, are due to their charge difference.\\

\begin{figure}[htbp]
    \begin{center}
        \includegraphics[height=4.5cm]{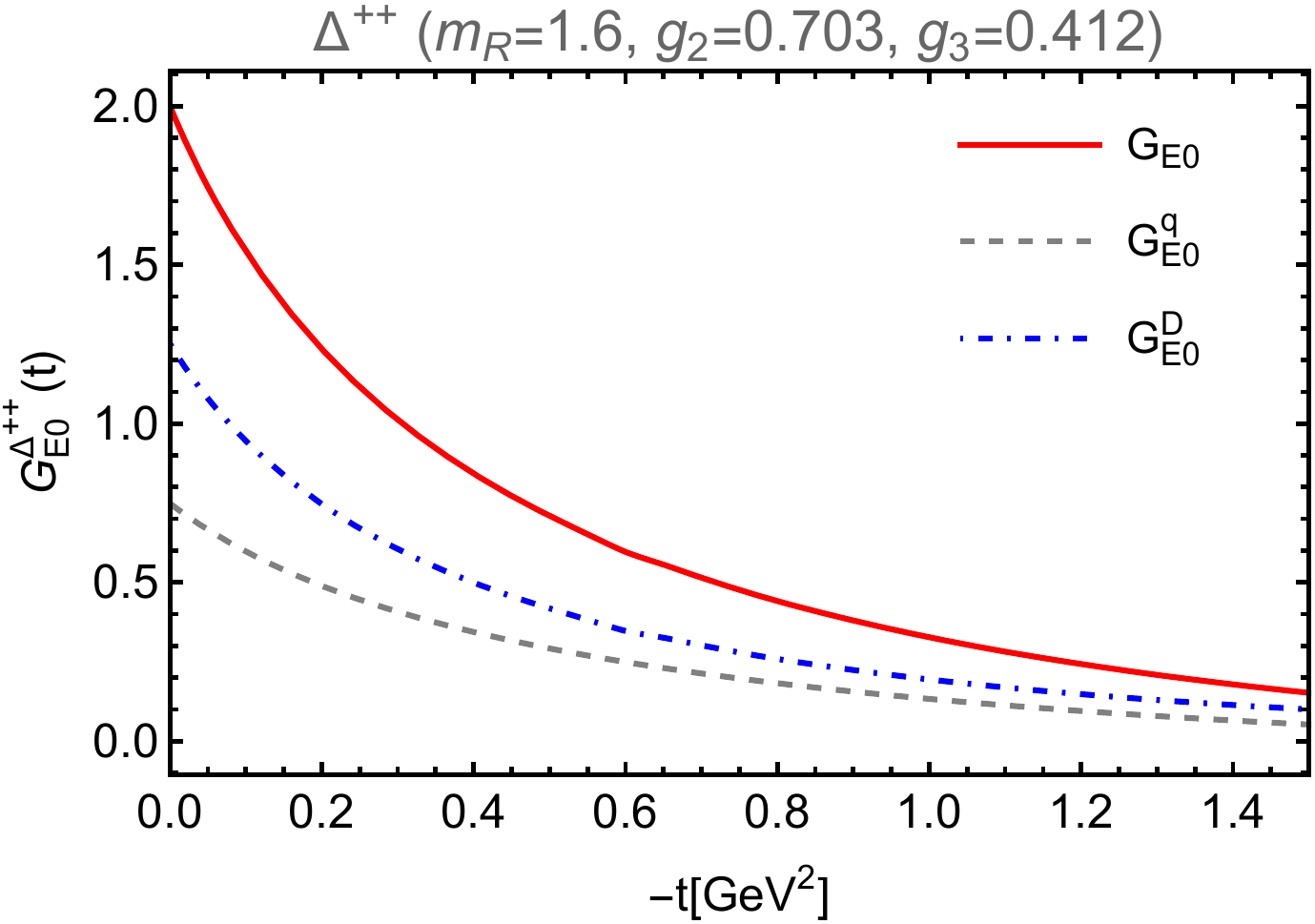}{\hskip 1.5cm}
        \includegraphics[height=4.5cm]{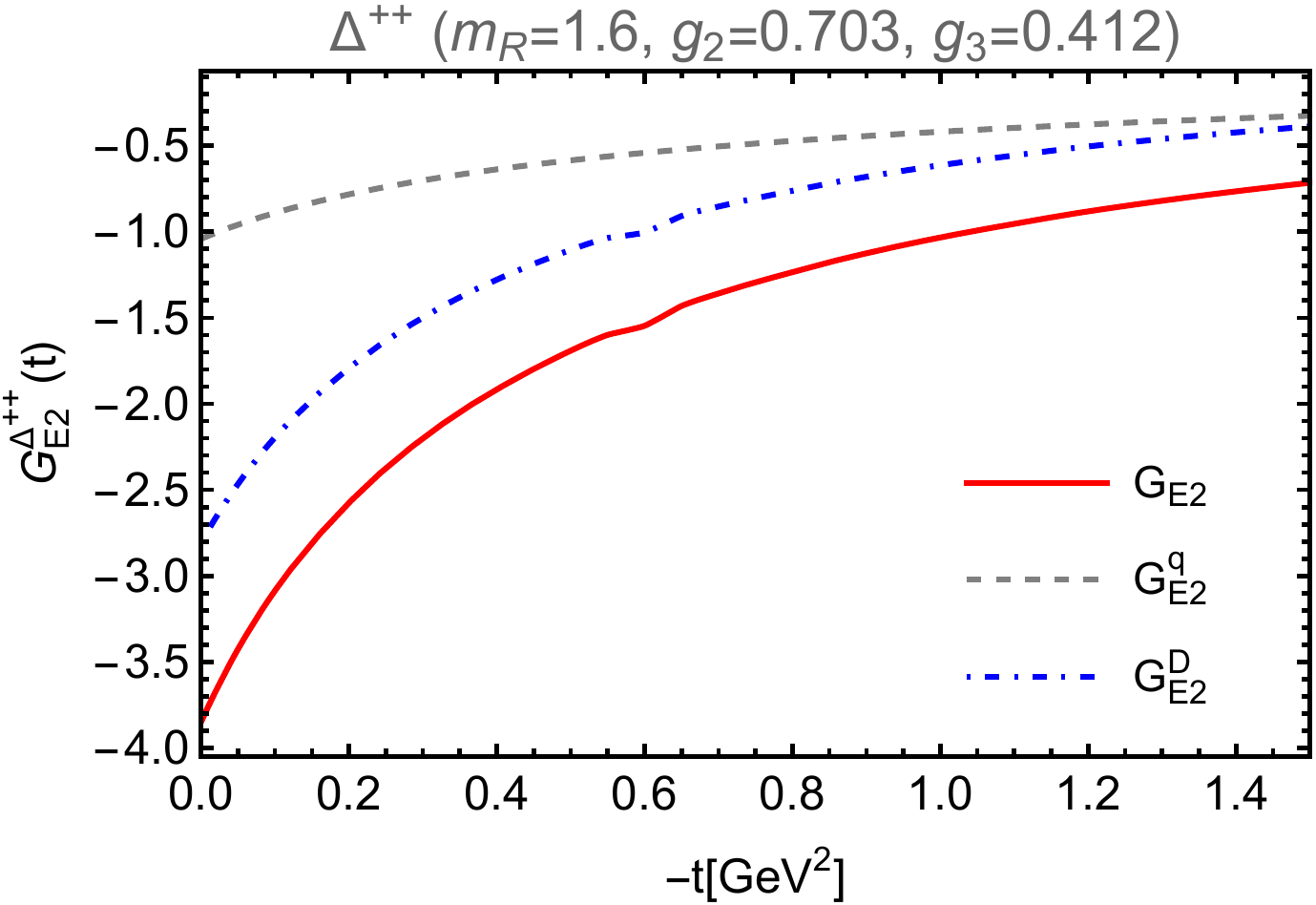}\\
        {\hskip 0.1cm}
        \includegraphics[height=4.48cm]{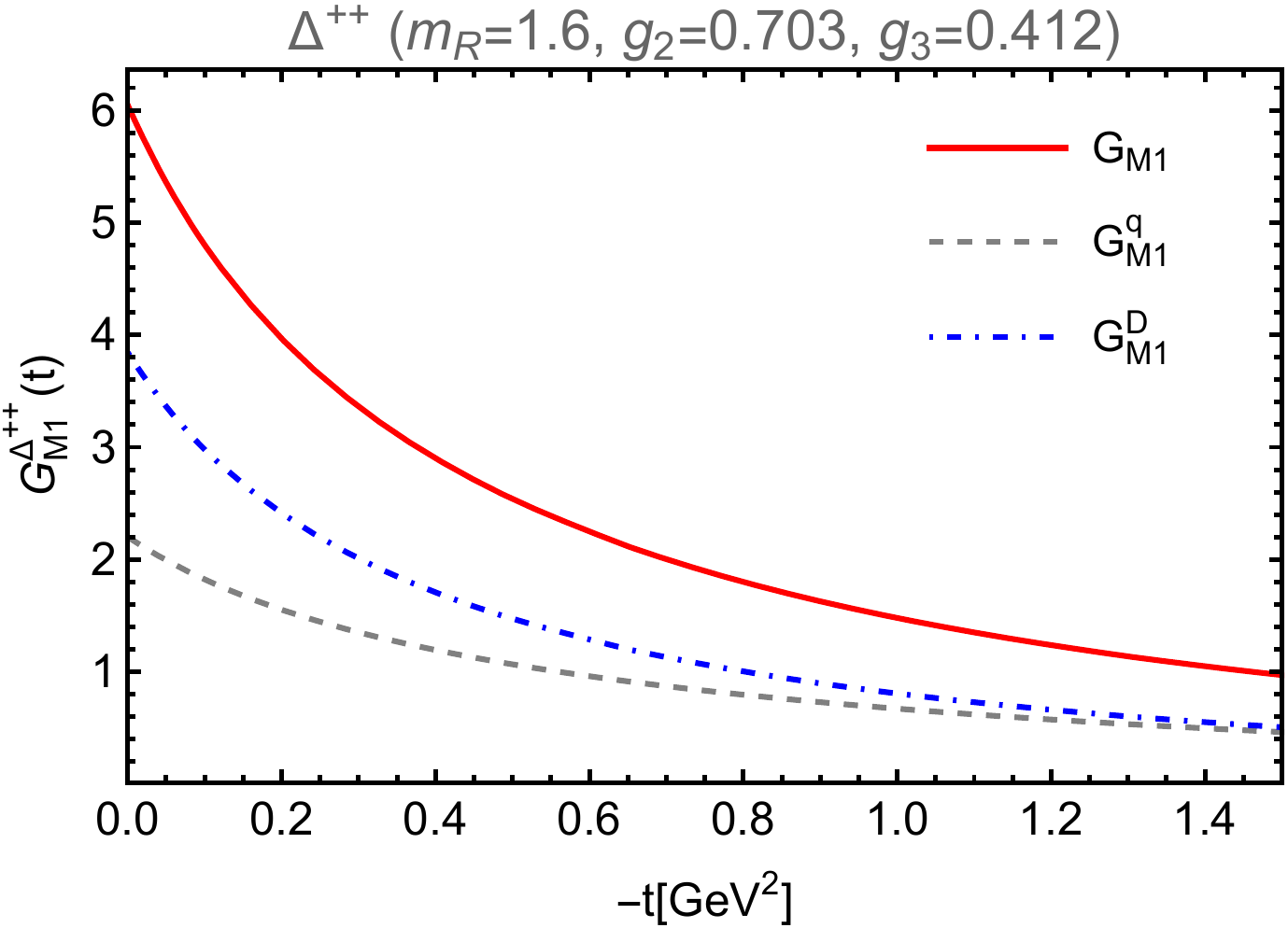}{\hskip 1.5cm}
        \includegraphics[height=4.5cm]{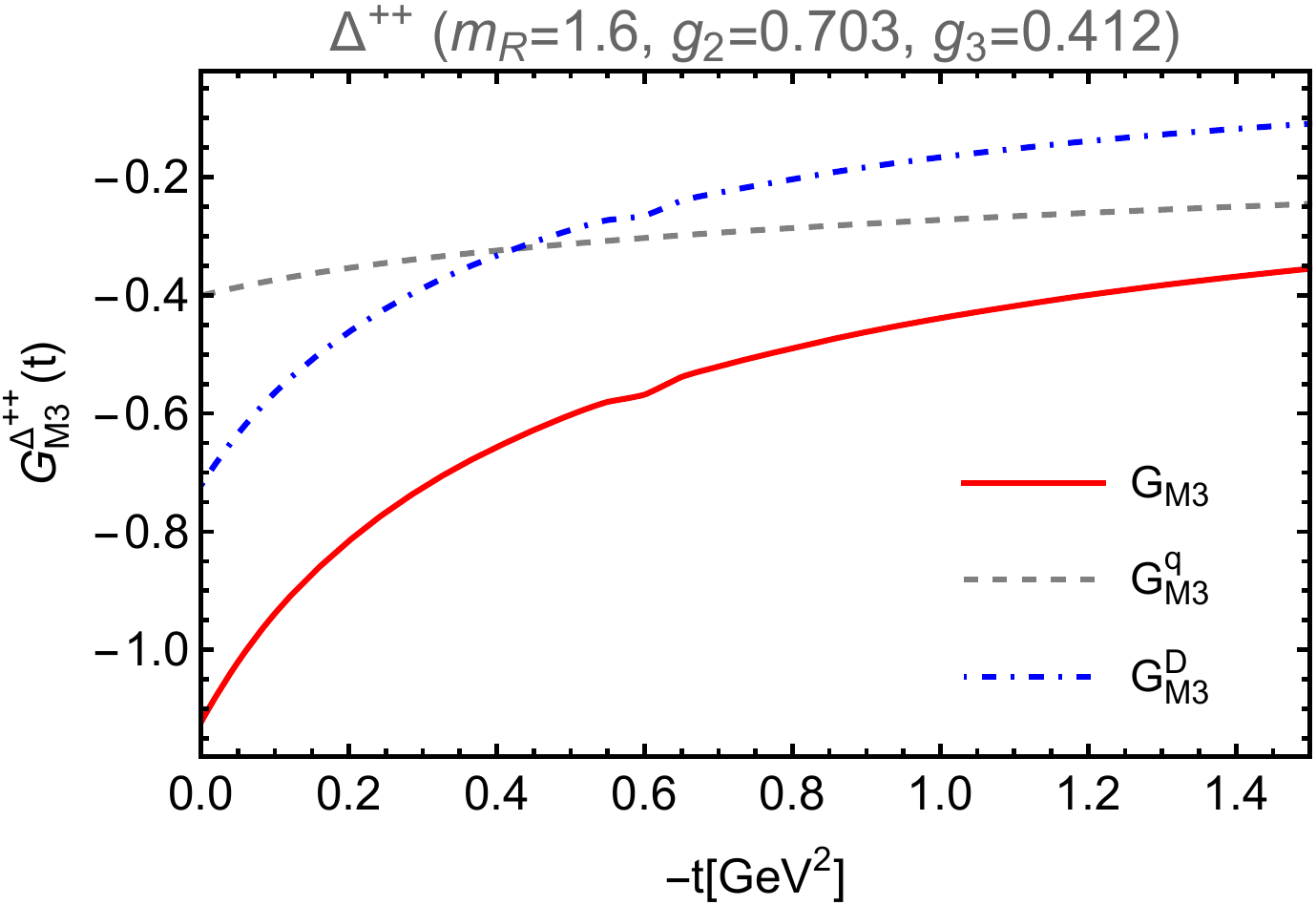}
        \caption{\small{The calculated four EMFFs of $\Delta^{++}$, the grey dashed, blue dashed-dotted, and red solid curves stand for the contributions from
        quark and diquark, and their sum, respectively.}}
        \label{EMFFuuu}
    \end{center}
\end{figure}

Figure~\ref{EMFFuuu} gives the individual contributions from the quark and diquark to the EMFFs of $\Delta^{++}$. As shown in the Figure, the ratio
of the contribution to EMFFs by diquark and quark is close to 2 as $-t$ goes to zero. It means that when the momentum
transfer is small the electromagnetic interaction probes the diquark as a point-like particle. This is consistent with the physical intuition and the
constituent quark model calculations~\cite{Krivoruchenko:1991pm,Capstick:1989ck,Capstick:1992uc,Capstick:1992xn}. However, when the momentum
transfer increases, the EM current probes much more inside the diquark such the effects of the binding and its quark structure become
remarkable. It should be addressed that, in the non-relativistic constituent quark model calculation, the coupling of each quark to the
electromagnetic probe is considered to be the same for simplicity and the total result is the three-times of the quark
contribution~\cite{Krivoruchenko:1991pm,Lipes:1972nf,Capstick:1989ck,Capstick:1992uc,Capstick:1992xn}, although the non-relativistic
wave function contains $\rho$ and $\lambda$ excitations. \\

Finally, we estimate the root mean squared (RMS)-radius of the $\Delta$ resonance according to our electric form factor $G_{E0}(t)$. It is
\begin{equation}
    \left\langle r^2 \right\rangle_E=0.665 \text{ fm}^2,
\end{equation}
for the charge distribution. It should be mentioned that the obtained charged RMS radii of
the three charged isospin partners $\Delta$ are the same since we do not consider the slight mass difference between the $u$ and $d$ quarks. \\

\subsection{The results of the matrix elements of EMT and GFFs of $\Delta$}

\subsubsection{EMT of $T^{00}$ and $T^{0i,i0}$}
\par\noindent\par
\begin{figure}[ht]
    \begin{center}
        {\hskip 0.01cm}
        \includegraphics[height=4.5cm]{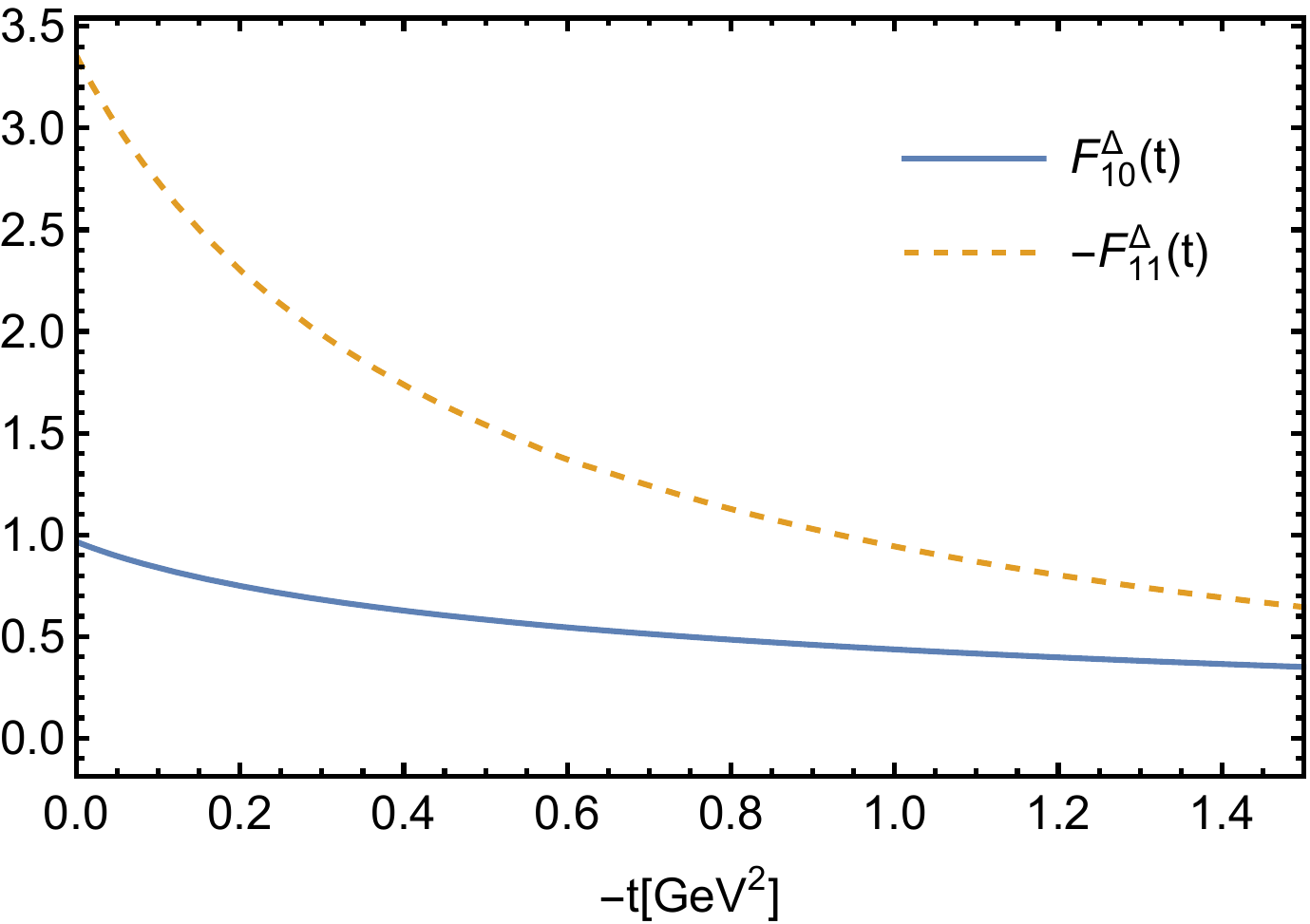}{\hskip 1.5cm}
        \includegraphics[height=4.5cm]{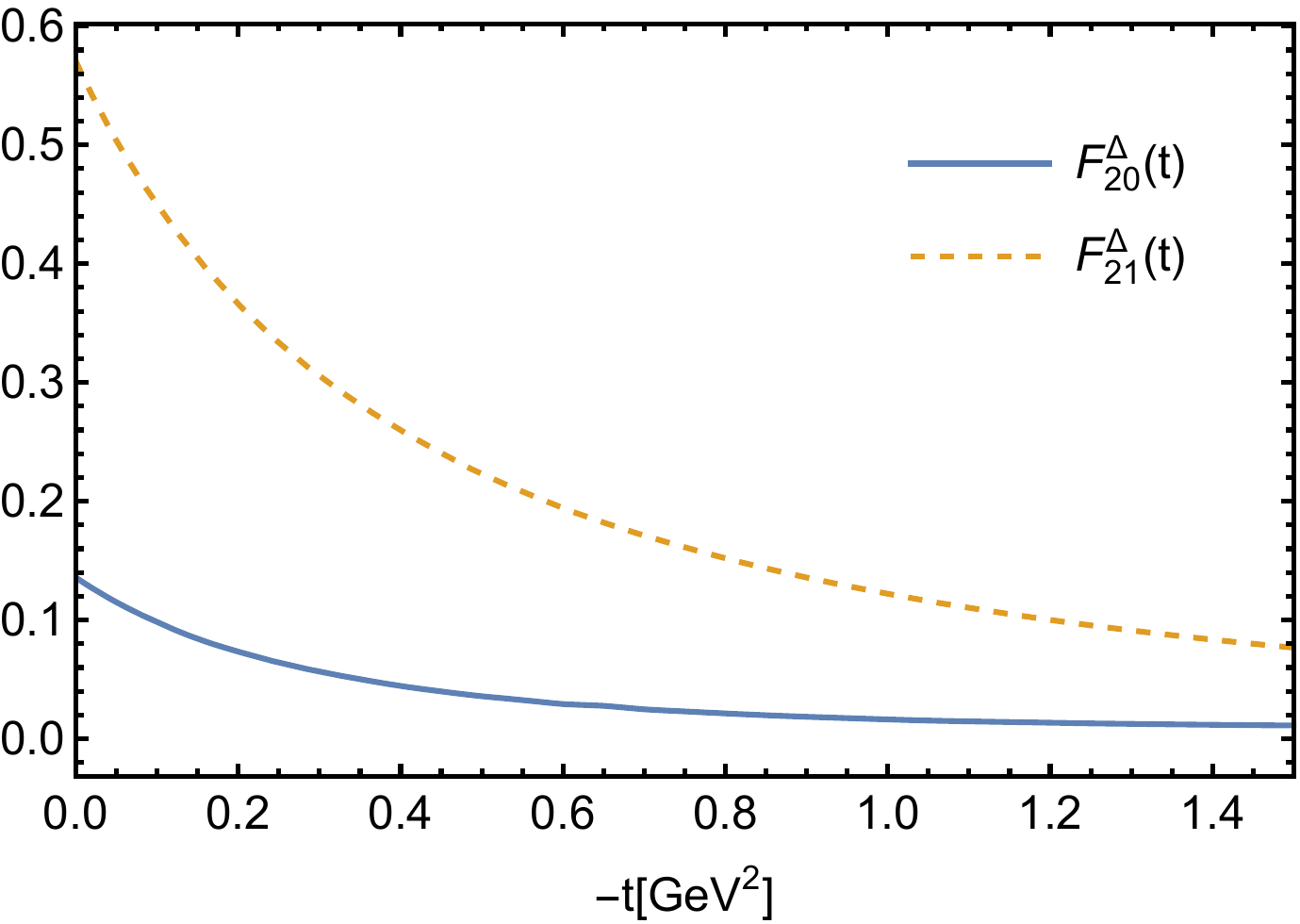}\\
        {\hskip 0cm}
        \includegraphics[height=4.42cm]{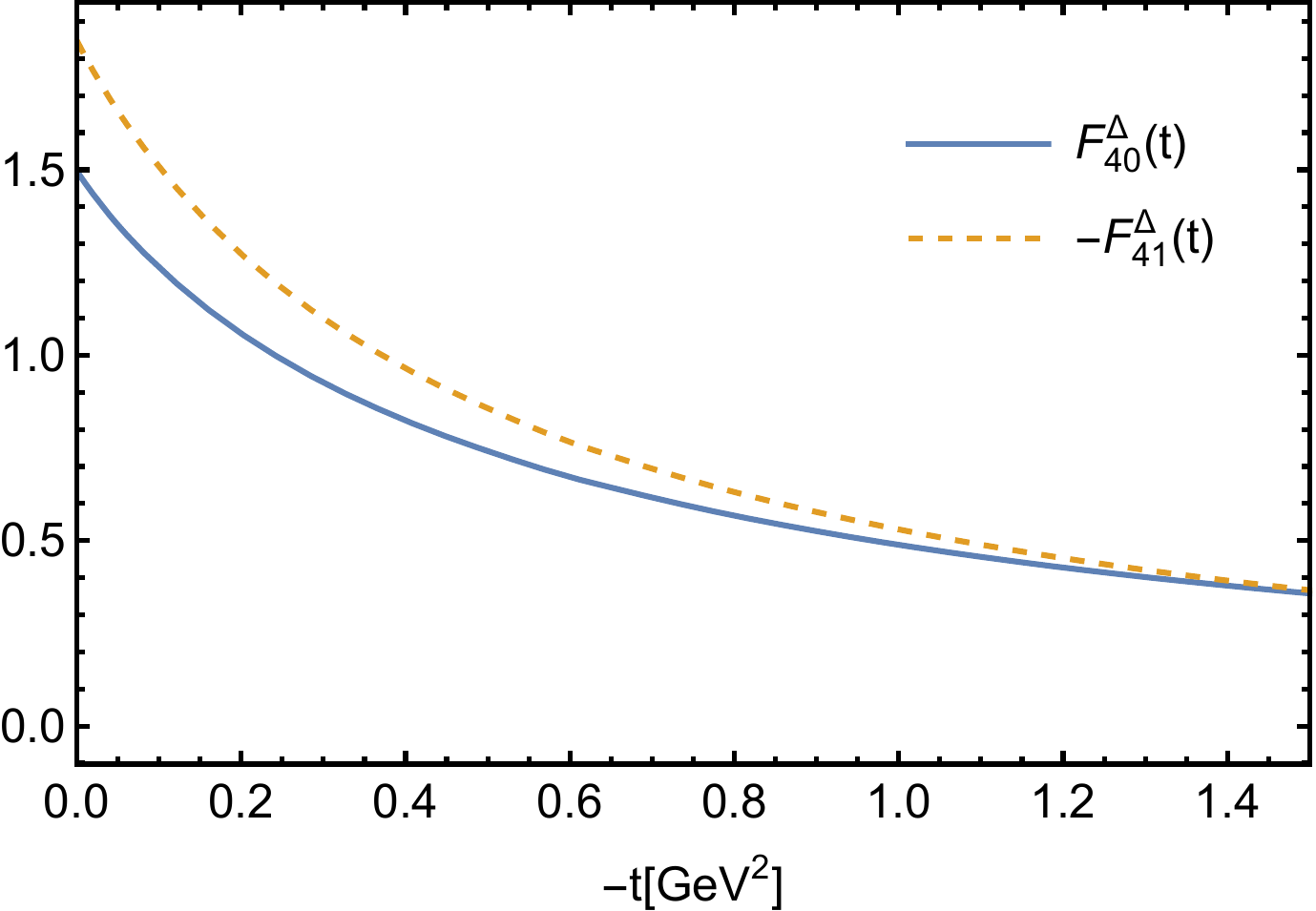}{\hskip 1.35cm}
        \includegraphics[height=4.42cm]{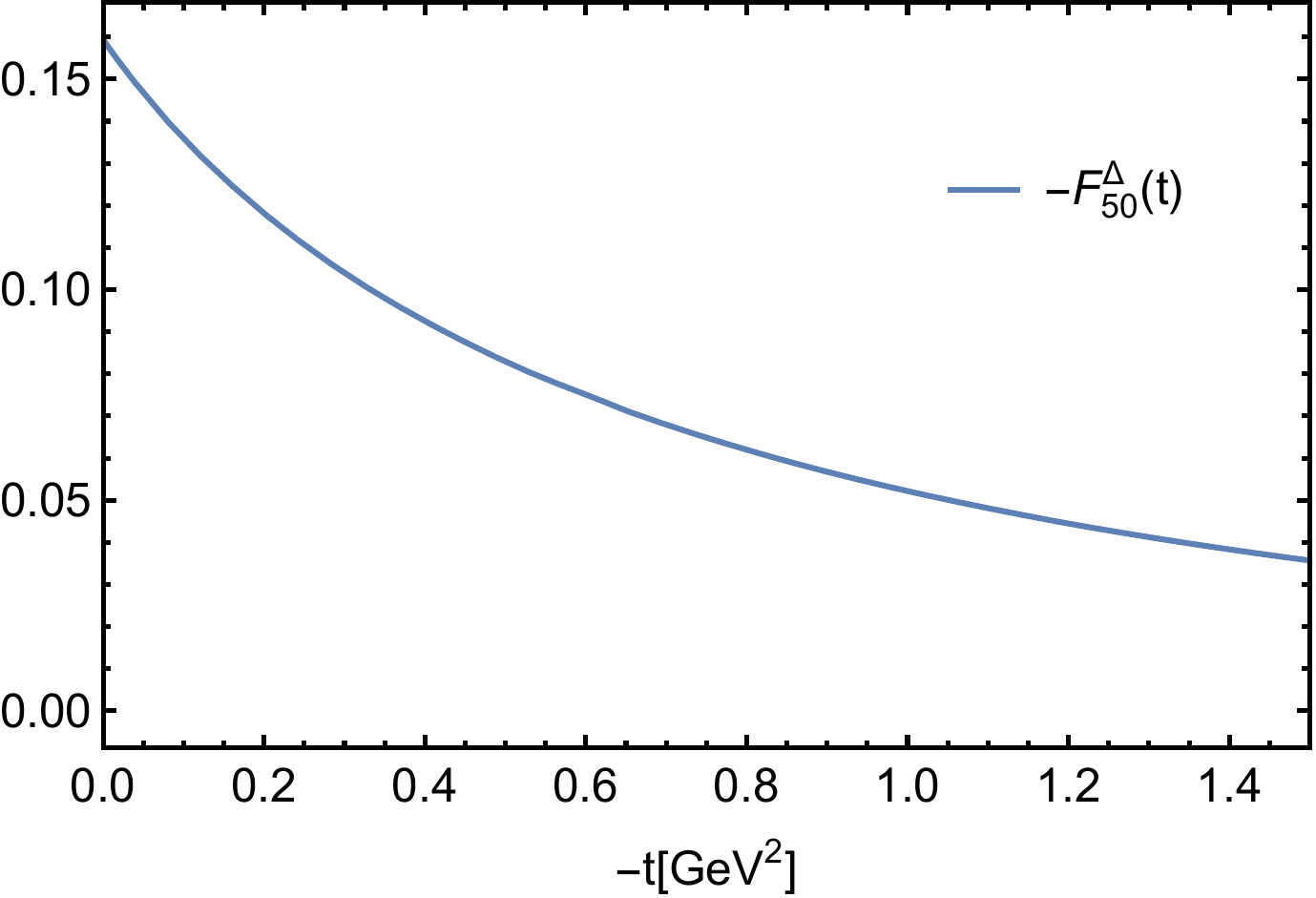}
        \caption{\small{Calculated GFFs of $F^T_{10,11,20,21,40,41,50}$ as functions of $-t$ for $\Delta$.}}
        \label{DeltaGMFFs1}
    \end{center}
\end{figure}
\begin{figure}[ht]
    \begin{center}
        \includegraphics[height=4.5cm]{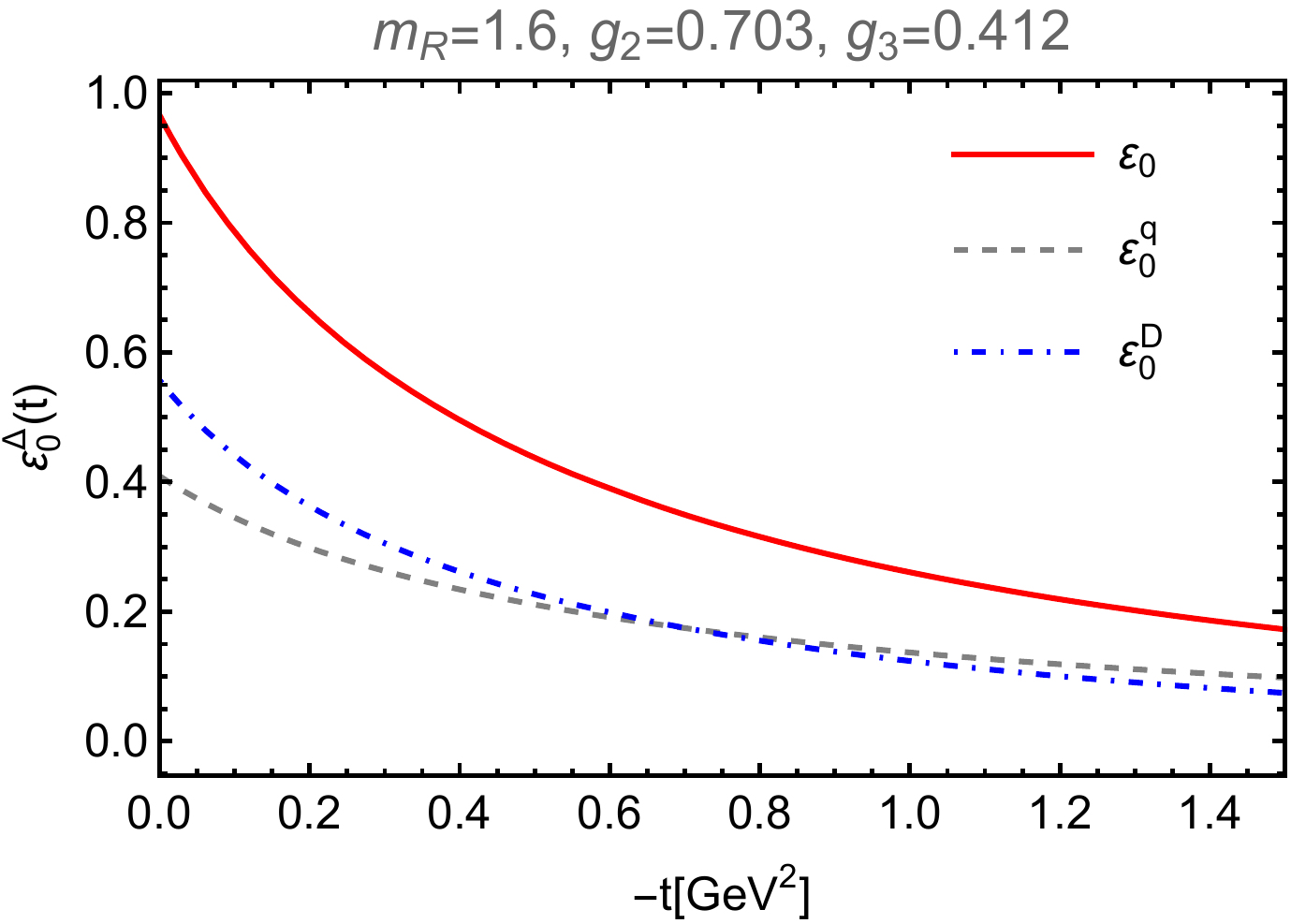}{\hskip 1.5cm}
        \includegraphics[height=4.5cm]{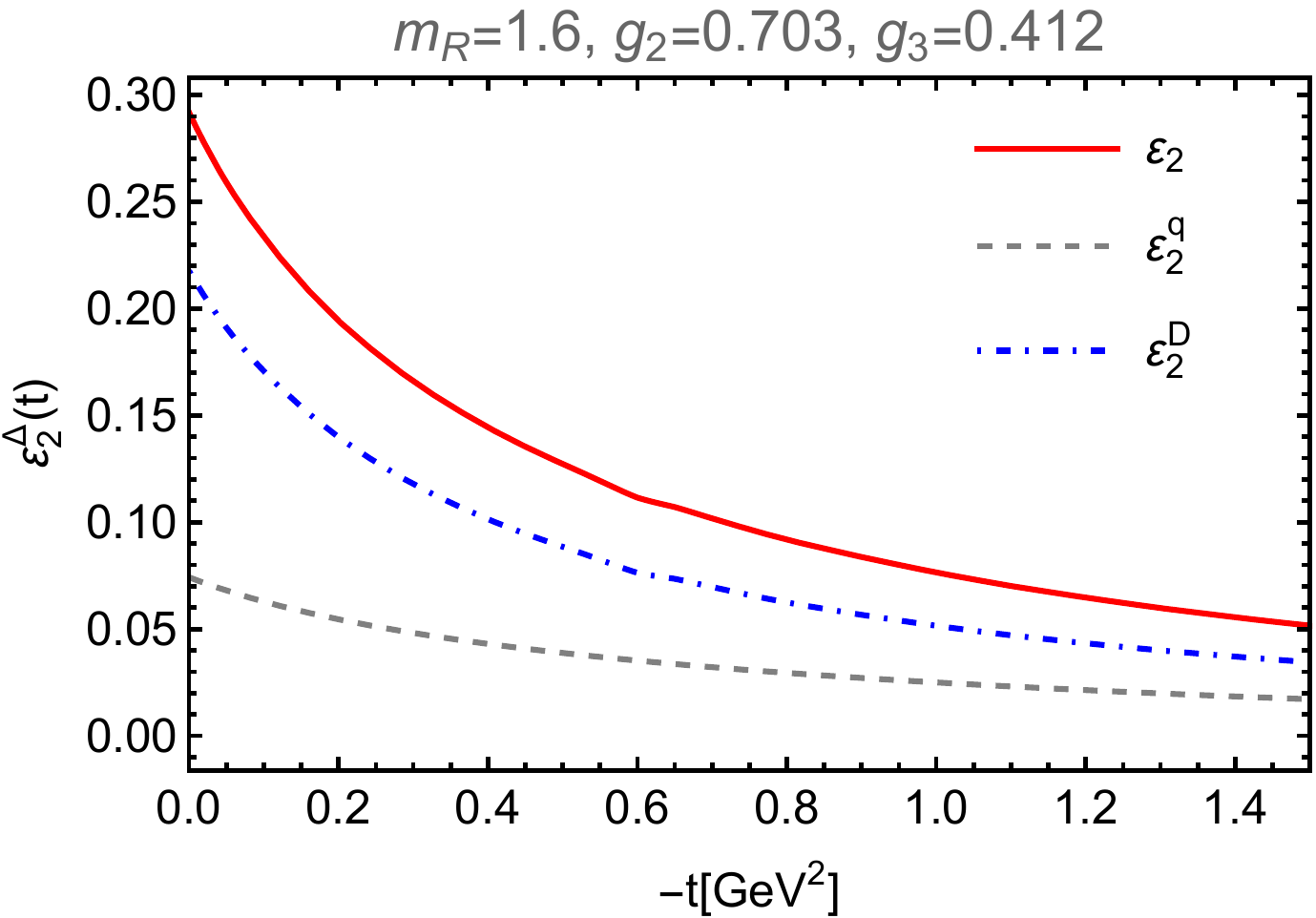}
        \caption{\small{The calculated energy monopole form factor of the $\Delta$ as a function of $-t$ (left panel) and the energy
         quadrupole (right panel). The dashed, dashed-dotted and solid curves stand for the contributions from quark, diquark and their sum.}}
        \label{GFFet}
    \end{center}
\end{figure}
\begin{figure}[h]
    \begin{center}
        \includegraphics[height=4.5cm]{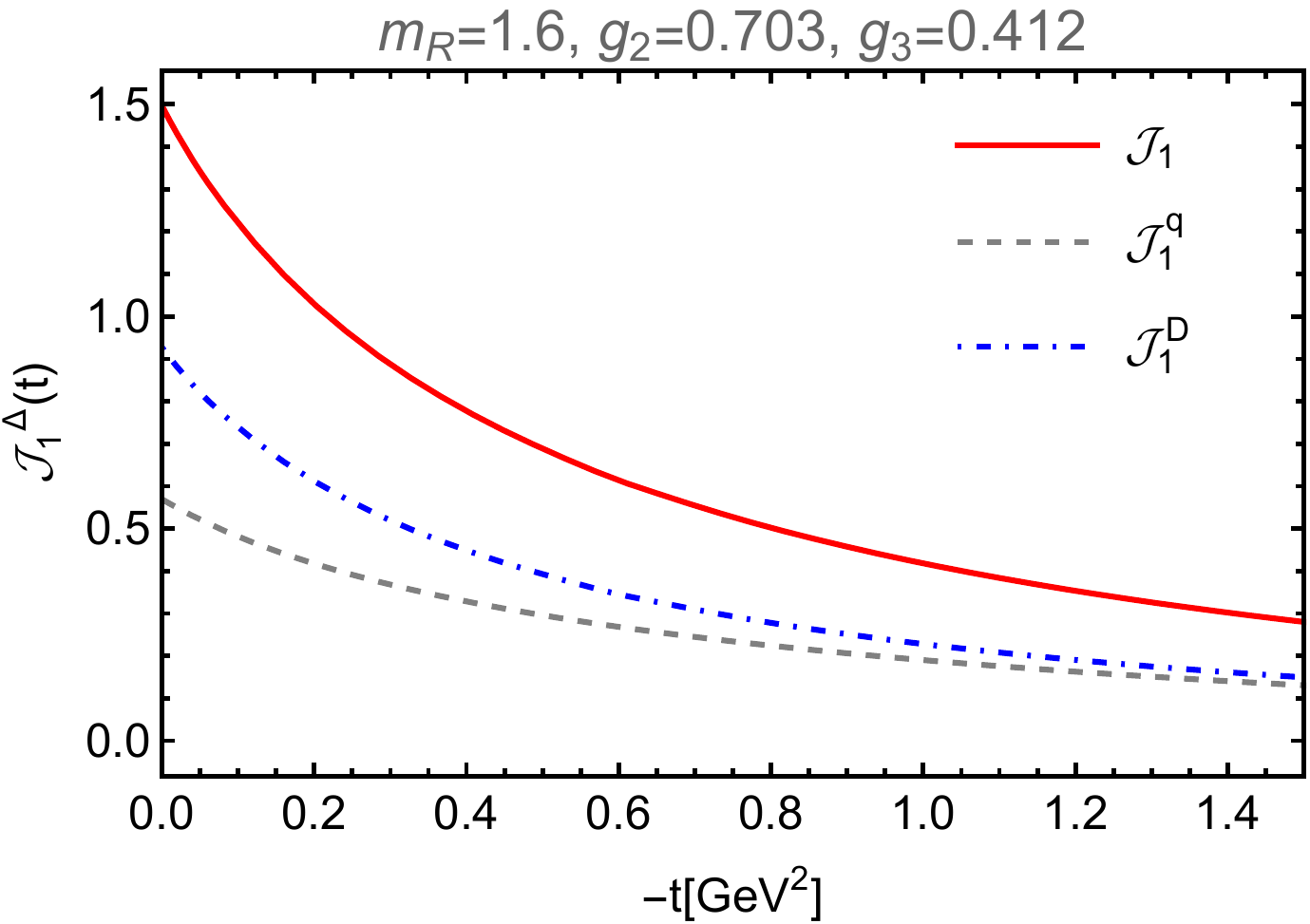}{\hskip 1.5cm}
        \includegraphics[height=4.5cm]{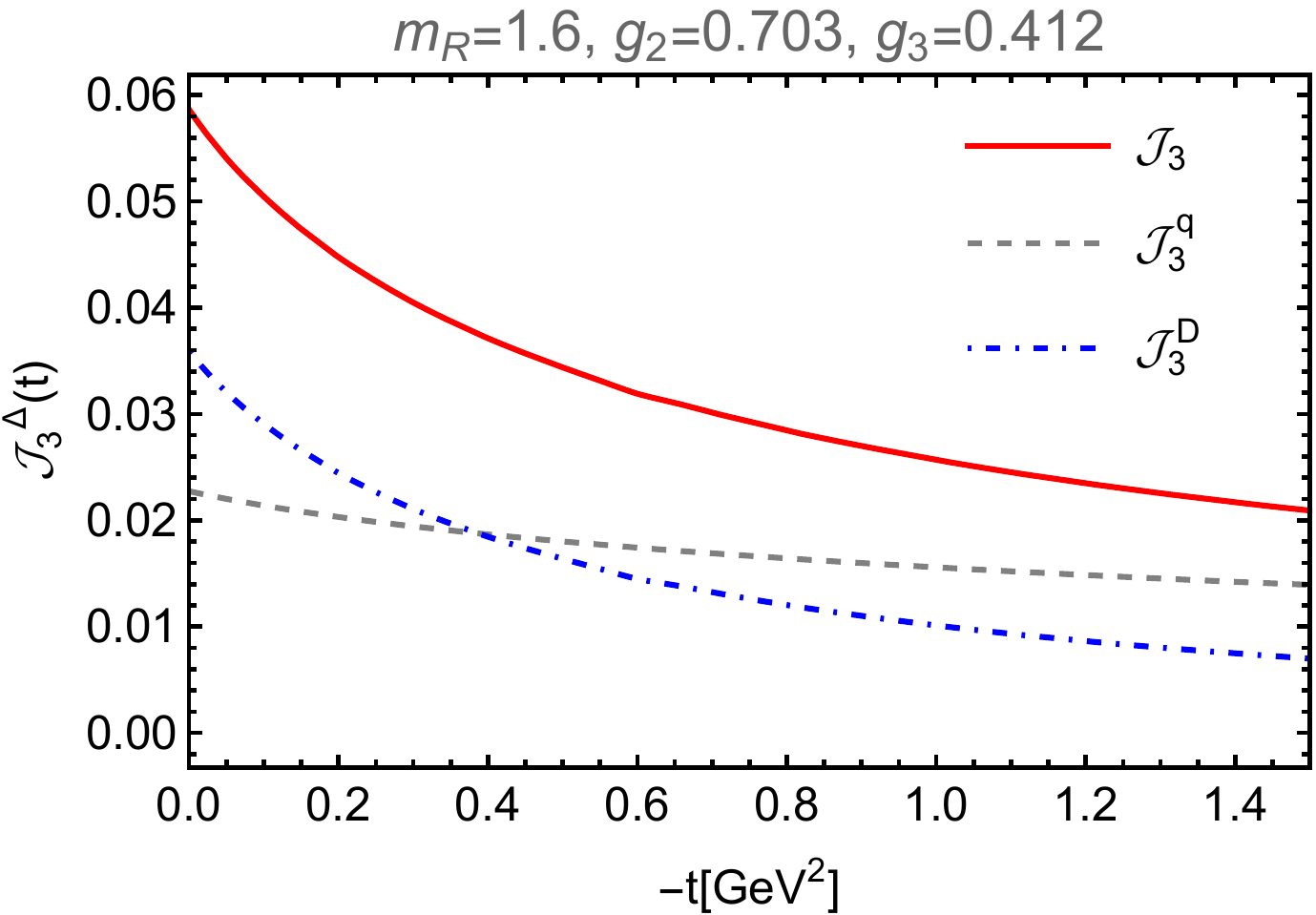}
        \caption{\small{The angular momentum form factor of the $\Delta$ as a function of $-t$ (left panel), and the octupole angular momentum form
        factor (the right panel). The solid, dashed and dashed-dotted curves represent the total result, and the contributions of quark and diquark,respectively.}}
        \label{GFFj}
    \end{center}
\end{figure}

Our relativistic covariant quark-diquark approach can be also applied for the calculations of the matrix element of the energy-momentum tensor
for the $\Delta$ spin-3/2 system according to the subsections~\ref{sub3_3} and \ref{sub3_4}. Here we show our results for GFFs as the functions of $-t$ in
Fig.~\ref{DeltaGMFFs1}, where the same normalization condition and model parameters are adopted as for the case of EMFFs.
By comparing our results with Ref.~\cite{Kim:2020lrs} where the Skyrme model is applied, we find that our $F^T_{1,(0,1)}$, $F^T_{4,(0,1)}$, and $F^T_{5,0}$ are consistent with
each other. However, our estimated $F^T_{2,(0,1)}$ have a big difference. This issue is closely related to the understanding of matrix
elements of $T^{ij}$ and it will be discussed later in detail. Then, we can reproduce the physical GMFFs from Eq.~\eqref{GMFFs_GFFs} for the energy and
angular momentum distributions. The results are displayed in Figs.~\ref{GFFet} and ~\ref{GFFj}.\\

As shown in Fig.~\ref{GFFet}, our $\varepsilon^{\Delta}_0(0) = 0.97\sim 1$ which correspond to the normalization condition of $\Delta$ mass. This result
indicates that the condition is not exactly preserved. This feature is expected to result from the off-shell effect due to the loop-integrals since our EMT
for free quark of Eq.~\eqref{Tquark} is conserved. The ratio of the contribution from diquark and quark is also close to 2 when $t=0$, similar to the
case of EMFFs.
Furthermore, as shown in Fig.~\ref{GFFj}, our estimated spin for $\Delta$ is $J^{\Delta}_1(0) \sim 1.5$ which just corresponds to the
total spin of $\Delta$ carried out by its two constituents.

The mass radius from Fig.~\ref{GFFet} is
\begin{equation}
    \left\langle r^2 \right\rangle_M=0.529 \text{ fm}^2,
\end{equation}
which is near but smaller than $\left\langle r^2 \right\rangle_E$. This number is close to $0.54~\text{fm}^2$ of Ref.~\cite{Kim:2020lrs}.\\

\begin{figure}[h]
    \begin{center}
        \includegraphics[height=4.5cm]{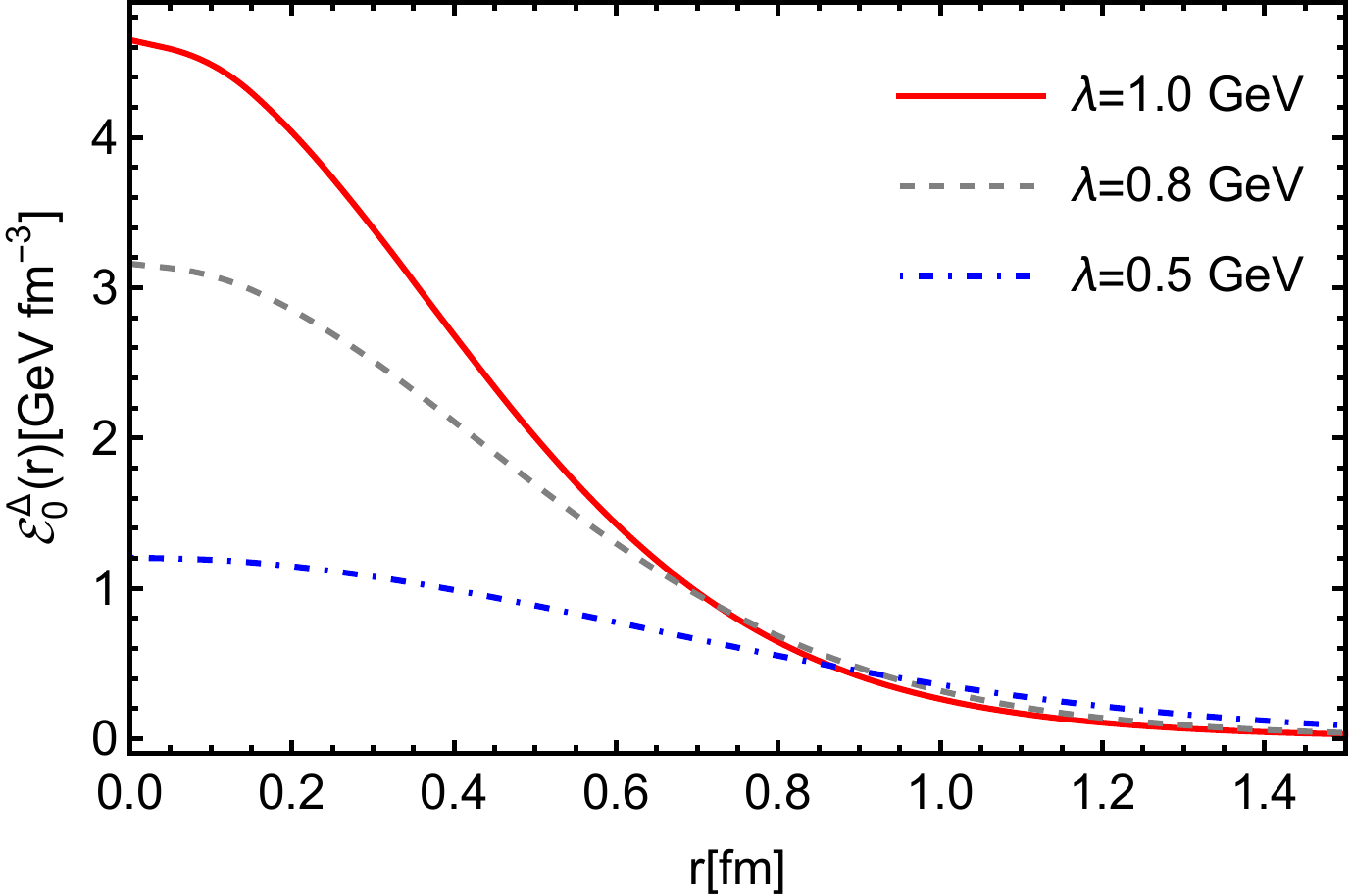}{\hskip 1.5cm}
        \includegraphics[height=4.5cm]{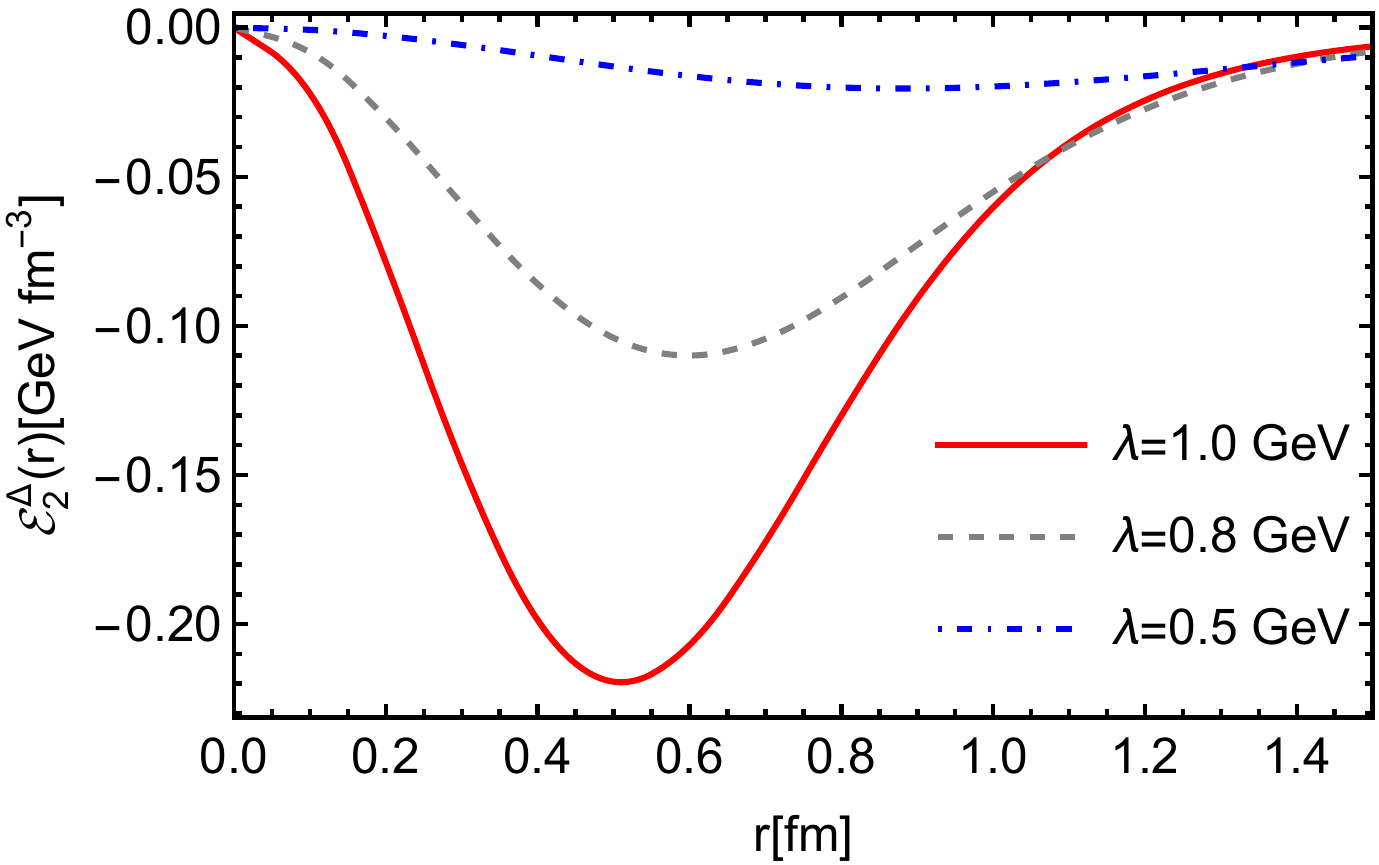}
        \caption{\small{The calculated energy monopole density of $\Delta$ as a function of $r$ (left panel) and
        energy quadrupole density (right panel). The Gaussian wave packet $e^{\frac{t}{\lambda^2}}$ has been included with
        $\lambda = 1~\text{GeV}$ (solid curve), $\lambda = 0.8~\text{GeV}$ (dashed curve), and $\lambda = 0.5~\text{GeV}$ (dashed-dotted curve).}}
        \label{GFFer}
    \end{center}
\end{figure}

Furthermore, the quantities, such as the energy densities and angular moment density can be obtained with the results given in Fig.~\ref{GFFer} by
performing the Fourier transformations as shown in Eq.~\eqref{epsi}. We know that the Fourier transformation of a plane wave is not well-defined,
and the transformations of our obtained GFFs, which are the functions of $-t$, cannot done due to the divergence. Thus, we add a Gaussian-like
wave packet $e^{\frac{t}{\lambda^2}}$~\cite{Meyer:2018twz,Sun:2018tmk} to guarantee the convergence when $|t|$ increases. Here, the model-parameter
$\lambda$ represents the size of the hadron with $\lambda\sim 1~\text{GeV}$. The inclusion of this additional factor is reasonable because of the locality
of the particle and the validity of the perturbative field theory. This issue has been discussed explicitly in Refs.~\cite{Diehl:2002he,Burkardt:2002hr}.
Our results for the densities in $r$-space are shown in Fig.~\ref{GFFer}. We find that the energy densities converge quickly to zero when $r > 1~\text{fm}$
and when $0.5~\text{GeV}<\lambda<1.1~\text{GeV}$.
Moreover, the Compton wavelength corresponding to $\lambda$ is about $2~\sim~4$ times the obtained radius of the $\Delta$ isobar.\\

\subsubsection{On the matrix elements of $T^{ij}$ and the $D$-term}
\par\noindent\par

It should be reiterated that our results for the GFFs of $F_{20}^\Delta(t)$ and $F_{21}^\Delta(t)$ shown in Fig.~\ref{DeltaGMFFs1} are different in sign
from the result of Ref.~\cite{Kim:2020lrs}. These two form factors and $F_{50}^\Delta(t)$ relate to the matrix element of $T^{ij}$ (see Eqs.~\eqref{eq:1} and \eqref{GMFFs_GFFs}).
In the classical mechanics of continuum media, the energy-momentum tensor $T^{ij}$ is interpreted as the pressure and shear force in the continuum
media approximation. When one discusses quantum field theory problems, in analogy to the classical mechanics for the continuum media, one expects that
the matrix element of $T^{ij}$ gives information of the pressure and shear force of the system. According to the relations between $T^{ij}$ and
the $D$-term and furthermore by considering the stability of the system, which implies that the corresponding pressure is positive, one concludes
that the $D$-term should be negative as $D_0(t=0)~<~0$ from Eq.~\eqref{Dtermps}.\\

Actually, the negativity of the $D$-term has been discussed extensively. Ref.~\cite{Perevalova:2016dln} explicitly proves this issue by discussing a
scalar hadron, which is assumed to be composed by two scalar fields. Under this circumstance, $D_0(t)$ is expressed as
(assuming the two constituents have same mass $m$, and the hadron has the mass $M=2m-B$ with $B$ being the binding energy)
\eq
\langle p'|T^{12}|p \rangle =\frac12 q^1 q^2D_0(t)=2ig^2
\int \frac{d^4k}{(2\pi)^4}\frac{(k^1-q^1/2)(k^2+q^2/2)+(k^2-q^1/2)(k^1+q^1/2)}{[(k-P)^2-m^2][(k+q/2)^2-m^2][(k-q/2)^2-m^2]},
\en
and
\eq
D_0(0)~\xlongequal[]{B\rightarrow 0}~-\frac{11}{3}+\frac{32}{3\pi}\sqrt{\frac{B}{2M}}-{\cal O}(\frac{B}{2M}),
\en
where the numerator in the first equation is due to the energy-momentum tensor of a scalar particle.
It is clearly seen that the expected $D$-term results from the sum of the numerators of $k^1k^2$ and $-q^1 q^2/4$. The first one
has a positive contribution while the second attributes a dominant negative value. Therefore, their sum gives $-\frac{11}{3}$.
When the binding $B$ increases, the calculated $D_0(0)$ reduces.\\

Inspired by the above analysis, the treatment of the hadrons, like $\Delta$ in this approach, is carried out in the following. Instead of considering two
spinless constituents, we take the fermion propagators, which is more realistic. Consequently, the matrix element of
$T^{\mu\nu}$ shown in Eqs.~\eqref{Dterm} and \eqref{DDterm} are much more complicated and much different from the one of scalar hadrons with
two scalar constituents.\\

In order to address the calculated matrix element of $T^{\mu\nu}$ more transparently and analytically, we simplify the Eq.~\eqref{Dterm} by
replacing the $\Gamma^{\alpha \beta}$ and $\Gamma^{\alpha' \beta'}$ with $c_1 g^{\alpha \beta}$ and $c_1 g^{\alpha' \beta'}$, and by replacing
the scalar function $[\left( l-P \right) ^2 -m_R^2+i \epsilon]^2 \biggl[\left( l-\frac{q}{2} \right) ^2 -m_R^2+i \epsilon\biggr]
\biggl[\left( l+\frac{q}{2} \right) ^2 -m_R^2+i \epsilon\biggr]$ in the denominator by $(l-P)^2 - m_R^2 + i \epsilon$.
And we omit $i \epsilon$ in writing for brevity.
We expect these replacements do not change the qualitative properties of our loop integrals. Then
\begin{equation}\label{DtermDetail}
	\begin{split}
		& \langle p', \lambda'\vert \hat{T}^{\mu \nu}_{q}(0) \vert p,\lambda\rangle\\
		&= - \bar{u}_{\alpha'}(p',\lambda') \frac{-i {\tilde C}^2}{2} \int \frac{d^4 l}{(2 \pi)^4}\frac{ g^{\alpha' \beta'}
\left( \slashed{l}+\frac{\slashed{q}}{2}+m_q \right)
		g_{\beta' \beta} (\gamma^\mu l^\nu + \gamma^\nu l^\mu ) \left( \slashed{l}-\frac{\slashed{q}}{2}
		+m_q \right) g^{\beta\alpha }}{[(l-P)^2-m_D^2][(l-\frac{q}{2})^2-m_q^2][(l+\frac{q}{2})^2-m_q^2][(l-P)^2-m_R^2]} u_{\alpha}(p,\lambda)\\
		& = - \bar{u}_{\alpha}(p',\lambda') \left(-i {\tilde C}^2\right) \int \frac{d^4 l}{(2 \pi)^4}
\frac{ {\tilde T}^{\mu \nu}}{[(l-P)^2-m_D^2][(l-\frac{q}{2})^2-m_q^2][(l+\frac{q}{2})^2-m_q^2][(l-P)^2-m_R^2]} u^{\alpha}(p,\lambda).
	\end{split}
\end{equation}

According to Eq.~\eqref{GFFsSpin3/2} and Appendix~C where $\TT^{\mu\nu}=\sum_i\TT^{\mu\nu}_i$, the $\frac{P^\mu P^\nu}{M}$ and $\frac{q^\mu q^\nu}{4M}$ terms
can be yielded by the standard Feynman parameterizations (see Appendix~B):
\begin{equation}
	\begin{split}
        & - i {\tilde C}^2 \int \frac{d^4 l}{(2 \pi)^4}\frac{1}{2M}\frac{\big [-l^2 \big (l^{\nu } P^{\mu }
+l^{\mu } P^{\nu}\big )+4 l^{\mu } l^{\nu }\left(l \cdot P\right) + 4m_qM l^{\mu } l^{\nu }+m_q^2 \big (l^{\nu } P^{\mu }
+l^{\mu } P^{\nu}\big )\big ]}{[(l-P)^2-m_D^2][(l-\frac{q}{2})^2-m_q^2][(l+\frac{q}{2})^2-m_q^2][(l-P)^2-m_R^2]}\\
		& = \frac{P^\mu P^\nu}{M} \tilde{\mathcal{A}} \int^1_0 dx_1 \int^{1-x_1}_0 dx_2 \int^{1-x_1-x_2}_0 dx_3 \frac{ M^2 (x_1+x_2)^3  + 2 m_q M (x_1+x_2)^2
+(\mathcal{M} + m_q^2)(x_1+x_2)}{ \mathcal{M}^2}\\
        & ~~~~ + \text{other~Lorentz~structures}\\
        & = \frac{P^\mu P^\nu}{M} F^T_{10}(0) + \text{other~Lorentz~structures},
    \end{split}
\end{equation}
and
\begin{equation}
	\begin{split}
& -i {\tilde C}^2 \int \frac{d^4 l}{(2 \pi)^4}\frac{2}{M}
\frac{  l^{\mu } l^{\nu } \left(l \cdot P\right)+  m_q Ml^{\mu } l^{\nu }
}{[(l-P)^2-m_D^2][(l-\frac{q}{2})^2-m_q^2][(l+\frac{q}{2})^2-m_q^2][(l-P)^2-m_R^2]}\\
& = \frac{q^\mu q^\nu}{4M} \tilde{\mathcal{A}} \int^1_0 dx_1 \int^{1-x_1}_0 dx_2 \int^{1-x_1-x_2}_0 dx_3
\frac{2[M^2 (x_1+x_2) + m_q M] (2x_3+x_1+x_2-1)^2}{\mathcal{M}^2}\\
&~~~~ + \text{other~Lorentz~structures} \\
& = \frac{q^\mu q^\nu}{4M}F_{20}^T(0) + \text{other~Lorentz~structures},
	\end{split}
\end{equation}
where
\eq
\tilde{\mathcal{A}}&=&\frac{ {\tilde C}^2 }{ (4\pi)^2 }~>~0,\\ \nonumber
\mathcal{M}&=&(x_1 + x_2)^2 M^2 - (x_1 + x_2) M^2 + x_1 m_D^2 + (1-x_1-x_2) m_q^2 + x_2 m_R^2~>~0.
\en
Moreover, $F_{50}^T (0) = 0$ in Eq.~\eqref{DtermDetail} for this simplified model.
Then our $\varepsilon_0(0) = F_{10}^T (0)$ and $D_0(0) = F_{20}^T (0)$, and both of the Feynman integrals are positive obviously. Therefore,
we conclude that the sign of the $D$ term in our calculation is the same as for $\varepsilon_0$. And the $D_0(0)~>~0$ can also be obtained from Eq.~\eqref{GMFFs_GFFs} and Fig.~\ref{DeltaGMFFs1} in our complete model,
and the von Laue condition~\cite{Laue:1911lrk} $\int^\infty_0 r^2 p(r) d r = 0$ is still satisfied.\\

We believe that our above conclusion is because of the Fermion properties and to the realistic consideration of the $\Delta$ isobar. This sign
problem also occurs in Ref.~\cite{Ji:2021mfb} when the hydrogen atom is considered. The controversial sign problem of the $D$-term is still
open. More realistic calculations for hadrons like nucleons are necessary to check if this problem indeed exists. It has been argued that the analogy to the
pressure in classical mechanics of $T^{ij}$ and the constraint of negativity of $D$-term may not be necessary. Instead, the momentum current
might be suitable to interpret the matrix element of $T^{ij}$ of a quantum system as argued in Ref.~\cite{Ji:2021mfb}.\\

\section{Summary and Conclusions}\label{section5}
\par\noindent\par

In this work we calculate the electromagnetic form factors and gravitational form factors of the spin-3/2 $\Delta$ with the
help of relativistic covariant quark-diquark approach. The internal quark structures of $\Delta$ as well as of the axial vector diquark are
explicitly considered. In order to simulate the bound state properties of $\Delta$ and diquark, we simply employ an ansatz for the vertex
scalar function, and the coupling of $\Delta$ to the quark and diquark, given by Ref.~\cite{Scadron:1968zz}, is adopted. We take
the Lattice QCD calculations for EMFFs as the constraints to fit our model-parameters.\\

It should be stressed that we simplify the three-body problem into a two-body problem by considering two quarks as a diquark.
To get more accurate results, we calculate the GFFs of the diquark as a two-body problem instead of just taking it as a point particle.
In section~\ref{section3}, we find that our results of EMFFs and electromagnetic moments are reasonable within acceptable region of $t$.
For the EMFFs of $\Delta$, the ratio of the contributions from the diquark and the quark is close to 2 when $t=0$. That is because they
are mainly determined by the number of charges, and the charge ratio of the diquark to the quark is 2. Similarly, because the mass ratio of
the diquark to the quark is close to 2, the contribution to GFFs from diquark is also close to 2 times of the corresponding one of
quark when $t=0$. Finally, we also reasonably reproduce the mass and spin distributions of $\Delta$. \\

However, we point out that there is a sign difference of our calculated $D$ term from the argument of its negativity. This is because of our realistic
consideration of the quark structures of $\Delta$ and of diquark, as we have shown in the detailed analyses of the matrix element of $T^{ij}$ and
of the Feynman loop integrals. It is argued that the $D$-term must be negative if a system satisfies the local stability criterion, otherwise if this was
not the case, the system would collapse~\cite{Perevalova:2016dln,Freese:2021mzg}. This argument originated from the interpretation of the stress tensor $T^{ij}$ as the momentum ﬂux and the normal force is expected to be outward. Our obtained positive $D$-term illustrates that its negativity might not be necessary. Instead, the momentum current interpretation
for the matrix element of $T^{ij}$ still might be suitable. More realistic studies for hadrons are needed to clarify this question. 
Finally, the present relativistic covariant quark-diquark approach will be employed for further studies of the GPDs of the $\Delta$ resonance and of
the $N-\Delta$ transition form factors. \\

\section*{Acknowledgements}
\par\noindent\par
We would like to thank Jambul Gegelia, Hyeondong Son, and June-Young Kim for valuable discussions and careful reading of the manucsript.
This work is supported by the National Natural Science Foundation of China under Grants Nos.~11975245, ~11947224, ~11947228, and~12035007. This work is also supported by the
State Scholarship Fund of China Scholarship Council under Grant No. 202006725011, the Sino-German CRC 110 ``Symmetries and the Emergence of Structure
in QCD'' project by NSFC under the Grant No.~12070131001, the Key Research Program of Frontier Sciences, CAS, under the Grant No.~Y7292610K1, and the
National Key Research and Development Program of China under Contracts No. 2020YFA0406300, and Guangdong Provincial funding with Grant No. 2019QN01X172.\\

\newpage
\normalem
\bibliographystyle{unsrt}
\bibliography{ref1}

\newpage
\setcounter{equation}{0}
\renewcommand\theequation{A.\arabic{equation}}
\section*{Appendix A: Some useful on shell identities}\label{usefulonshell}
\par\noindent\par
To compute the matrix element of EMT current and electromagnetic current, some identities explicitly given in Ref.~\cite{Lorce:2017isp} are employed.
These identities are satisfied for the Rarita-Schwinger spinors. In terms of the variables $P=(p'+p)/2$ and $q=p'-p$,
\begin{equation}\label{pd}
    P^\alpha \doteq\frac{q ^\alpha}{2},\qquad
    P^{\alpha '} \doteq-\frac{q ^{\alpha '}}{2},
\end{equation}
where $\doteq$ means on-shell equality, and we reserve the indices $\alpha_i$ and $\alpha_i'$. There are some on-shell relations derived from the
Gordon identity and the Schouten identity,
\begin{equation}
    \bar{u}(p',\lambda')\gamma^\mu u(p,\lambda)=
    \bar{u}(p',\lambda ') \biggl[ \frac{P^\mu}{M} +\frac{i \sigma^ {\mu \nu}
    q_\nu}{2M} \biggr] u(p,\lambda),
\end{equation}
\begin{equation}
    i \epsilon^{\mu \nu \rho \sigma} g^{\tau \lambda} + i \epsilon^{\nu \rho \sigma \tau} g^{\mu \lambda} + i \epsilon^{\rho \sigma \tau \mu} g^{\nu \lambda} + i \epsilon^{\sigma \tau \mu \nu} g^{\rho \lambda} + i \epsilon^{\tau \mu \nu \rho} g^{\sigma \lambda} = 0.
\end{equation}
We can rewrite the Gordon identity using on shell equality
\begin{equation}\label{gordon}
    \gamma^\mu \doteq \frac{P^\mu}{M} +\frac{i \sigma^ {\mu \nu}
    q_\nu}{2M}.
\end{equation}
The other on-shell relations used in our work read~\cite{Lorce:2017isp}
\begin{mysubeq}
    \begin{align}
        \mathtt{1} &\doteq \frac{\slashed{P}}{M}, & \qquad 0 &\doteq \slashed{q},\\
        \gamma_5 &\doteq\frac{\slashed{q} \gamma_5}{2M}, &\qquad
        0 &\doteq\slashed{P}\gamma_5,\\
        \gamma^\mu &\doteq \frac{P^\mu}{M} + \frac{i \sigma ^{\mu q}}{2M},
        &\qquad 0 &\doteq \frac{q ^ \mu}{2} + i \sigma ^ {\mu P},\\
        \gamma^\mu \gamma_5 &\doteq \frac{q^\mu \gamma_5}{2M}
        + \frac{i \sigma ^ {\mu P}}{M}, &\qquad 0 &\doteq P^\mu \gamma_5
        + \frac{i \sigma ^{\mu q} \gamma_5}{2},\\
        i \sigma^{\mu \nu} &\doteq - \frac{q ^{[ \mu} \gamma^{\nu ]}}{2M}
        +\frac{i \epsilon^{\mu \nu P \lambda} \gamma_\lambda \gamma_5}{M},
        &\qquad 0 &\doteq -P^{[ \mu} \gamma ^{\nu ]}
        + \frac{i \epsilon^{\mu \nu q \lambda} \gamma_\lambda \gamma_5}{2},\\
        i \sigma^{\mu \nu} \gamma_5 &\doteq -\frac{P^{[ \mu} \gamma ^{\nu ]} \gamma_5}{M}
        + \frac{i \epsilon^{\mu \nu q \lambda} \gamma_\lambda}{2M},
        &\qquad 0 &\doteq -\frac{q ^{[ \mu} \gamma^{\nu ]} \gamma_5}{2}
        + i \epsilon^{\mu \nu P \lambda} \gamma_\lambda,
    \end{align}
\end{mysubeq}
where $\sigma ^ {\mu P} \equiv \sigma ^ {\mu \nu} P_\nu$, $\epsilon^{\mu \nu P \lambda} \equiv
\epsilon^{\mu \nu \rho \lambda} P_\rho $.\\

The Rarita-Schwinger spinors satisfy this relation,
\begin{equation}\label{rs}
    \gamma^{\alpha_i} u_{\alpha_1 \dots \alpha_n} \left( p, \lambda \right) =0, \qquad
    \bar{u}_{\alpha'_1 \dots \alpha'_n} \left( p',\lambda' \right) \gamma^{\alpha'_i}=0,
    \qquad i ~ \in ~ \{ 1,\dots ,n \}.
\end{equation}
Combining Eqs.~\eqref{pd}, \eqref{gordon} and \eqref{rs}, we can get these on-shell identities,
\begin{equation}
    i \sigma ^{\alpha' \mu} \doteq g ^ {\alpha' \mu},
    \qquad i \sigma ^ {\nu \alpha} \doteq g ^ {\nu \alpha}.
\end{equation}
Some important on-shell identities we used are derived from the product of three and four Dirac matrices,
\begin{mysubeq}\label{three}
    \begin{align}
    \gamma^\rho \gamma^\mu \gamma^\sigma & = g ^{\rho \mu} \gamma ^\sigma -g ^{\rho \sigma}
    \gamma^\mu + g ^{\mu \sigma} \gamma^\rho - i \epsilon ^{\rho \mu \sigma \lambda}
    \gamma_\lambda \gamma_5, \\
    \gamma^\rho \gamma^\mu \gamma^\sigma \gamma_5 & =  g ^{\rho \mu} \gamma ^\sigma \gamma_5
    -g ^{\rho \sigma}\gamma^\mu \gamma_5 + g ^{\mu \sigma} \gamma^\rho \gamma_5
    - i \epsilon ^{\rho \mu \sigma \lambda} \gamma_\lambda,
    \end{align}
\end{mysubeq}
\begin{equation}\label{four}
    \begin{split}
    \gamma^\rho \gamma^\mu \gamma^\nu \gamma^\sigma = & \mbox{ }g^{\rho \mu} g^{\nu \sigma}
    -g^{\rho \nu} g^{\mu \sigma} + g^{\rho \sigma} g^{\mu \nu}
    + i \epsilon^{\rho \mu \nu \sigma} \gamma_5 - g ^{\rho \mu} i \sigma^{\nu \sigma}\\
    &+ g^{\rho \nu} i \sigma ^{\mu \sigma} - g^{\rho \sigma} i \sigma ^ {\mu \nu}
    -g^{\nu \sigma} i \sigma^{\rho \mu} + g^{\mu \sigma} i \sigma^{\rho \nu}
    -g^{\mu \nu} i \sigma ^{\rho \sigma}.
    \end{split}
\end{equation}

The nontrivial relation obtained using Eqs.~\eqref{gordon}, \eqref{three}, and \eqref{four}~\cite{Nozawa:1990gt},
\begin{equation}
    q^{\alpha'} g^{\mu \alpha} - q^\alpha g^{\mu \alpha'} \doteq 2M \left(
    1-\frac{q^2}{4M^2} \right) g^{\alpha' \alpha} \gamma^\mu - 2 g^{\alpha' \alpha}
    P^\mu + \frac{1}{M} q^{\alpha'} q^\alpha \gamma^\mu.
\end{equation}
Because of Eqs.~\eqref{gordon} and \eqref{pd} this identity can be derived~\cite{Cotogno:2019vjb},
\begin{equation}\label{jian}
    \frac{q^2}{2} q^{[ \alpha'} g^{\alpha ] [ \mu} P^{\nu ]} \doteq
    - q^{\alpha'} q^\alpha P^{[ \mu} i \sigma^{\nu ] q}
    + P^2 q^{[ \alpha'} g^{\alpha ] [ \mu} i \sigma^{\nu ] q}.
\end{equation}
Another nontrivial relation was derived from Ref.~\cite{Cotogno:2019vjb},
\begin{equation}\label{jia}
    \begin{split}
    & q^2 g^{\mu \nu} g^{\alpha' \alpha} - 2 g^{\mu \nu} q^{\alpha'} q^\alpha
    - g^{\alpha' \alpha} P^{\{ \mu} i \sigma^{\nu \} q} + q^{[ \alpha'}
    g^{\alpha ] \{ \mu} P^{\nu \}} - g^{\alpha' \alpha} q^\mu q^\nu +
    \frac{1}{2} q^{\{ \alpha'} g^{\alpha \} \{ \mu} q^{\nu \}} \\
    &~~~\doteq \frac{1}{2} q^{[ \alpha'} g^{\alpha ] \{ \mu} i \sigma^{\nu \} q}
    - \frac{1}{2} q^{\{ \alpha'} g^{\alpha \} \{ \mu} q^{\nu \} }
    +q^2 g^{\alpha' \{ \mu} g^{\nu \} \alpha}.
    \end{split}
\end{equation}
And combining Eqs.~\eqref{jian} and \eqref{jia}, we can obtain
\begin{equation}
    \begin{split}
    q^{[ \alpha'} g^{\alpha ] \mu} i \sigma^{\nu q}
    \doteq & q^2 g^{\mu \nu} g^{\alpha' \alpha} - 2 g^{\mu \nu} q^{\alpha'} q^\alpha - g^{\alpha' \alpha} P^{\{ \mu} i \sigma^{\nu \} q} + q^{[ \alpha'}
    g^{\alpha ] \{ \mu} P^{\nu \}} - g^{\alpha' \alpha} q^\mu q^\nu \\
    & + q^{\{ \alpha'} g^{\alpha \} \{ \mu} q^{\nu \}} - q^2 g^{\alpha' \{ \mu} g^{\nu \} \alpha} + \frac{q^2}{4P^2} q^{[ \alpha'} g^{\alpha ]
    [ \mu} P^{\nu ]} + \frac{1}{2P^2} q^{\alpha'} q^\alpha P^{[ \mu} i \sigma^{\nu ] q},
    \end{split}
\end{equation}
\begin{equation}
    \begin{split}
    q^{[ \alpha'} g^{\alpha ] \nu} i \sigma^{\mu q}
    \doteq &  q^2 g^{\mu \nu} g^{\alpha' \alpha} - 2 g^{\mu \nu} q^{\alpha'} q^\alpha - g^{\alpha' \alpha} P^{\{ \mu} i \sigma^{\nu \} q} + q^{[ \alpha'}
    g^{\alpha ] \{ \mu} P^{\nu \}} - g^{\alpha' \alpha} q^\mu q^\nu \\
    & + q^{\{ \alpha'} g^{\alpha \} \{ \mu} q^{\nu \}} - q^2 g^{\alpha' \{ \mu} g^{\nu \} \alpha} - \frac{q^2}{4P^2} q^{[ \alpha'} g^{\alpha ]
    [ \mu} P^{\nu ]} - \frac{1}{2P^2} q^{\alpha'}q^\alpha P^{[ \mu} i \sigma^{\nu ] q}.
    \end{split}
\end{equation}
There are still some more identities, see Ref.~\cite{Cotogno:2019vjb}.

\setcounter{equation}{0}
\renewcommand\theequation{B.\arabic{equation}}
\section*{Appendix B: Feynman parameterization and Loop integrals}\label{feynmanp}
\par\noindent\par
In our calculation, we use the Feynman parameterization. Some integrals are listed as follows.
\begin{mysubeq}\label{loren}
 \begin{align}
    -i \int \frac{d^4 l}{(2 \pi)^4}\frac{1}{\mathfrak{D}}=&A_{00},\\
    -i \int \frac{d^4 l}{(2 \pi)^4}\frac{l^ \mu}{\mathfrak{D}}=&A_{11} P^\mu,\\
    -i \int \frac{d^4 l}{(2 \pi)^4}\frac{l^ \mu l^\nu}{\mathfrak{D}}=&A_{21} g^{\mu \nu}
    +A_{22} P^\mu P^\nu +A_{23} q^\mu q^\nu,\\
    -i \int \frac{d^4 l}{(2 \pi)^4}\frac{l^ \mu l^\nu l^\gamma}{\mathfrak{D}}=
    &A_{31} \left(P^\gamma g^{\mu  \nu } + P^\mu g^{\gamma  \nu} + P^\nu g^{\gamma  \mu}\right)
    + A_{32} \left(q ^\mu q ^\nu P^\gamma + q^\gamma q ^\nu P^\mu
    + q^\gamma q^\mu P^\nu\right)\nonumber \\
    &~~+A_{33} P^\gamma P^\mu P^\nu,\\
    -i \int \frac{d^4 l}{(2 \pi)^4}\frac{l^ \mu l^\nu l^\gamma l^\rho}{\mathfrak{D}}=
    &\mathop {\sum}_{\underset{\in (\mu ,\nu , \gamma,\rho)}{(i,j,m,n)}}
    \left[ \frac{1}{8} A_{41}g^{i j } g^{m n}+\frac{1}{8}A_{42} P^i P^j g^{m n}
    +\frac{1}{8} A_{43} q^i q^j g^{m n} \right.\nonumber \\
    &~~~\left.+\frac{1}{24} A_{44} P^i P^j P^m P^n  +\frac{1}{24} A_{45} q^i q^j q^m q^n
    +\frac{1}{8}A_{46} q^i q^j P^m P^n \right],
     \end{align}
\end{mysubeq}
\noindent{\hskip -0.3cm}
where $A_{n_1n_2}$ stand for the structural integrals. The symmetric properties of the denominator $\mathfrak{D}$ with respect to $q$, as
shown in Eq.~\eqref{DeltaD}, is considered.\\
\vspace{1cm}

\setcounter{equation}{0}
\renewcommand\theequation{C.\arabic{equation}}
\section*{Appendix C: Calculation details about $D$-term}\label{detaild}
\par\noindent\par
\appendix
According to Eq.~\eqref{DtermDetail}
\begin{equation}
	\begin{split}
		& \langle p', \lambda'\vert \hat{T}^{\mu \nu}_{q}(0) \vert p,\lambda\rangle\\
		&= - \bar{u}_{\alpha'}(p',\lambda') \frac{-i {\tilde C}^2}{2} \int \frac{d^4 l}{(2 \pi)^4}\frac{ g^{\alpha' \beta'}
\left( \slashed{l}+\frac{\slashed{q}}{2}+m_q \right)
		g_{\beta \beta'} \gamma^{ \{ \mu} l^{\nu\} } \left( \slashed{l}-\frac{\slashed{q}}{2}
		+m_q \right) g^{\alpha \beta}}{[(l-P)^2-m_D^2][(l-\frac{q}{2})^2-m_q^2][(l+\frac{q}{2})^2-m_q^2][(l-P)^2-m_R^2]} u_{\alpha}(p,\lambda)\\
		& = - \bar{u}_{\alpha}(p',\lambda') \left(-i {\tilde C}^2\right) \int \frac{d^4 l}{(2 \pi)^4}
\frac{ {\tilde T}^{\mu \nu}}{[(l-P)^2-m_D^2][(l-\frac{q}{2})^2-m_q^2][(l+\frac{q}{2})^2-m_q^2][(l-P)^2-m_R^2]} u^{\alpha}(p,\lambda),
	\end{split}
\end{equation}
where
\begin{equation}
    {\tilde T}^{\mu \nu}= \frac{1}{2} \left( \slashed{l}+\frac{\slashed{q}}{2}+m_q \right) \gamma^{ \{ \mu} l^{ \nu \} }
     \left( \slashed{l}-\frac{\slashed{q}}{2} + m_q \right).
\end{equation}
Here ${\tilde T}^{\mu \nu}$ contains the information of Fermions and can be divided into nine parts $\tilde T^{\mu \nu} =\overset{9}{\underset{i=1}{\sum}}
\tilde T_i^{\mu \nu}$. Using the on-shell identities given in Appendix~A, we get
\begin{mysubeq}
\begin{align}
		\TT_1^{\mu \nu} =& ~\frac{1}{2}\slashed{l} \gamma^{ \{ \mu} l^{ \nu \} } \slashed{l}
\doteq \frac{i l^{\mu } l^{\nu } \sigma ^{l q } }{M}-\frac{i l^2 l^{\{\mu } \sigma ^{\nu\} q} }{4 M} - \frac{l^2 l^{ \{\mu } P^{\nu \} } }{2 M}
+ \frac{2 l^{\mu } l^{\nu } \left(l \cdot P\right)}{M},\\ \nonumber
		\TT_2^{\mu \nu} =& ~\frac{1}{4}\slashed{q} \gamma^{ \{ \mu} l^{ \nu \} } \slashed{l} \doteq \frac{i q^{ \{ \mu } l^{\nu \} }
\sigma ^{l q} }{8 M} - \frac{i l^{ \{ \mu } \sigma ^{ \nu \} q }  \left(l \cdot  q\right)}{8 M} - \frac{i l^{ \{ \mu } P^{\nu \} } \sigma ^{l q} }{4 M}
        +\frac{i l^{ \{ \mu } \sigma ^{ \nu \} q}  \left(l \cdot   P\right)}{4 M}\\
& +\frac{q^{ \{ \mu } l^{\nu \} }  \left(l \cdot   P\right)}{4 M}-\frac{l^{ \{ \mu } P^{\nu \} }  \left(l \cdot  q\right)}{4 M},\\
		\TT_3^{\mu \nu} =& ~\frac{1}{2}m_q \gamma^{ \{ \mu} l^{ \nu \} } \slashed{l} \doteq -\frac{1}{2} i m_q l^{ \{ \mu } \sigma ^{\nu \}  l}
 +m_q l^{\mu } l^{\nu } ,\\
		\TT_4^{\mu \nu} =& -\frac{1}{4} \slashed{l} \gamma^{ \{ \mu} l^{ \nu \} } \slashed{q} \doteq
-\frac{i q^{ \{ \mu } l^{\nu \} } \sigma ^{l q} }{8 M}+\frac{i l^{ \{ \mu } \sigma ^{ \nu \} q }
\left(l \cdot  q\right)}{8 M} -\frac{i l^{ \{ \mu } P^{\nu \} } \sigma ^{l q} }{4 M}
        +\frac{i l^{ \{ \mu } \sigma ^{\nu \} q }  \left(l \cdot   P\right)}{4 M}\\ \nonumber
&  + \frac{l^{ \{ \mu } P^{\nu \} }  \left(l \cdot  q\right)}{4 M}-\frac{q^{ \{ \mu } l^{\nu \} }  \left(l \cdot   P\right)}{4 M},\\
		\TT_5^{\mu \nu} =& -\frac{1}{8} \slashed{q} \gamma^{ \{ \mu} l^{ \nu \} } \slashed{q} \doteq
        \frac{i q^2 l^{ \{ \mu } \sigma ^{ \nu \} q } }{16 M}+\frac{q^2 l^{ \{ \mu } P^{\nu \} } }{8 M},\\
		\TT_6^{\mu \nu} =& - \frac{1}{4} m_q  \gamma^{ \{ \mu} l^{ \nu \} } \slashed{q} \doteq
        \frac{1}{4} i m_q l^{ \{ \mu } \sigma ^{ \nu \} q } -\frac{1}{4} m_q q^{ \{ \mu } l^{\nu \} } ,\\
		\TT_7^{\mu \nu} =& ~\frac{1}{2} m_q \slashed{l} \gamma^{ \{ \mu} l^{ \nu \} }  \doteq
        \frac{1}{2} i m_q l^{ \{ \mu } \sigma ^{ \nu \} l } +m_q l^{\mu } l^{\nu } ,\\
		\TT_8^{\mu \nu} =& ~\frac{1}{4} m_q \slashed{q} \gamma^{ \{ \mu} l^{ \nu \} }  \doteq
        \frac{1}{4} i m_q l^{ \{ \mu } \sigma ^{ \nu \} q } +\frac{1}{4} m_q q^{ \{ \mu } l^{\nu \} } ,\\
		\TT_9^{\mu \nu} =& ~\frac{1}{2}m_q^2 \gamma^{ \{ \mu} l^{ \nu \} }  \doteq
        \frac{i m_q^2 l^{ \{ \mu } \sigma ^{ \nu \} q } }{4 M}+\frac{m_q^2 l^{ \{ \mu } P^{\nu \} } }{2 M}.
\end{align}
\end{mysubeq}

According to Appendix~B, we see that the $\frac{P^\mu P^\nu}{M}$ term comes from the loop integrals of $\TT_1^{\mu \nu}$, $\TT_3^{\mu \nu}$, $\TT_7^{\mu \nu}$,
and $\TT_9^{\mu \nu}$, and the $\frac{q^\mu q^\nu}{4M}$ term, which contributes to $D$-term, results from the ones of $\TT_1^{\mu \nu}$, $\TT_3^{\mu \nu}$,
and  $\TT_7^{\mu \nu}$.
\end{document}